\documentclass{IEEEtran}

\usepackage{acronym} 

\usepackage{cite}

\ifCLASSINFOpdf
  \usepackage[pdftex]{graphicx}
  \graphicspath{./imgs/}
  \DeclareGraphicsExtensions{.pdf,.jpeg,.png}
\else
  \usepackage[dvips]{graphicx}
  \graphicspath{./imgs/}
  \DeclareGraphicsExtensions{.eps}
\fi

\usepackage{epstopdf}
\usepackage{tikz}
\usepackage{url}
\usepackage[cmex10]{amsmath}
\usepackage{mathtools} 
\usepackage{color}
\usepackage{cleveref}

\usepackage{cite}

\usepackage[export]{adjustbox}

\ifCLASSOPTIONcompsoc
  \usepackage[caption=false,font=normalsize,labelfont=sf,textfont=sf,justification=centering]{subfig}
\else
  \usepackage[caption=false,font=footnotesize,justification=centering]{subfig}
\fi
\usepackage{stfloats}
\renewcommand*{\Im}{\operatorname{Im}}

\usepackage{algorithm}
\usepackage{algpseudocode}
\usepackage{amsfonts}
\usepackage{float}
\usepackage{multirow}

\usepackage[colorinlistoftodos]{todonotes}

\algnewcommand\algorithmicinput{\textbf{Input:}}
\algnewcommand\Input{\item[\algorithmicinput]}
\algnewcommand\algorithmicoutput{\textbf{Output:}}
\algnewcommand\Output{\item[\algorithmicoutput]}

\begin{document}
%
\title{Non-Stationary Power System Forced Oscillation Analysis using Synchrosqueezing Transform}
%
%
%

\author{Pablo Gill Estevez, Pablo Marchi, Cecilia Galarza and Marcelo Elizondo 
\thanks{P. Gill Estevez, P. Marchi, and C. G. Galarza are with the School of Engineering, Universidad de Buenos Aires and the CSC-CONICET, Argentina. M. Elizondo works with Pacific Northwest National Laboratory, Seattle, USA. (e-mail: pgill@fi.uba.ar, pmarchi@fi.uba.ar, cgalar@fi.uba.ar, marcelo.elizondo@pnnl.gov)}}%
\maketitle

\begin{abstract}
Non-stationary forced oscillations (FOs) have been observed in power system operations. However, most detection methods assume that the frequency of FOs is stationary. In this paper, we present a  methodology for the analysis of non-stationary FOs. Firstly, Fourier synchrosqueezing transform (FSST) is used to provide a concentrated time-frequency representation of the signals that allows identification and retrieval of non-stationary signal components. To continue, the Dissipating Energy Flow (DEF) method is applied to the extracted components to locate the source of forced oscillations. The methodology is tested using simulated as well as real PMU data. The results show that the proposed FSST-based signal decomposition provides a systematic framework for the application of DEF Method to non-stationary FOs.
\end{abstract}

\begin{IEEEkeywords}
Forced oscillations, non-stationary signal, phasor measurement unit (PMU), time-frequency analysis, synchrosqueezing, multicomponent signals. 
\end{IEEEkeywords}

%
\IEEEpeerreviewmaketitle

\section{Introduction}
%
%
%
%

\IEEEPARstart{U}{nlike} modal oscillations, which mainly depend on the dynamic characteristics of the system, forced oscillations (FOs) are determined by inputs and disturbances that drive the system \cite{Follum2016, Follum2017}. FOs can occur in power systems due to different causes, such as equipment failure, inadequate control designs, and abnormal generator operating conditions \cite{Nerc2017}. The sustained presence of significant forced oscillations on the power system could lead to long-term adverse effects. For example, equipment fatigue and possible damage to rotor shafts or power quality reduction. It is clear that monitoring these oscillations, understanding how they act on power systems, and implementing mitigation strategies are relevant matters to be considered \cite{Nerc2017}.

The most efficient way for mitigating sustained oscillations is to locate the source and to disconnect it from the network. This action requires locating the system component causing the oscillations \cite{Chen2013}. Many methods for locating the source have been proposed in the past few years, and each of them has advantages and disadvantages and can be successfully used only for specific circumstances \cite{Dan2018}. Among these methods, the Dissipating Energy Flow (DEF) method \cite{Chen2013} has shown the best performance, and it was recently applied to data obtained from a practical experience \cite{MASLENNIKOV201755}. Most methods for analyzing FOs assume that the frequency of the oscillation source is stationary \cite{Follum2016,Nerc2017,Chen2013,MASLENNIKOV201755,Zhou2015,Sarmadi2016,GHORBANIPARVAR2017,Jha2019}. However, past events have shown that FO fundamental frequency could be non-stationary (for example, in the October 3, 2017 event in ISO-NE) \cite{Maslennikov2018}. 

Synchrosqueezing Transform (SST) is a time-frequency (TF) analysis technique that was designed to decompose signals into constituent components with time-varying oscillatory characteristics \cite{DAUBECHIES2011243}. SST is an alternative to the Empirical Mode Decomposition (EMD) method \cite{Huang1998} with a stronger analytical foundation \cite{Hemakom2017}. SST was originally introduced in the context of Continuous Wavelet Transform (CWT) \cite{DAUBECHIES2011243}. The univariate CWT-based SST (WSST) reassigns the wavelet coefficients in scale or frequency by combining the coefficients that contain the same instantaneous frequencies (IF), such that the resulting energy is concentrated around the IF curves of the modulated oscillations. A natural extension of WSST was proposed in \cite{Oberlin2014} by using the short-time Fourier transform (STFT). This technique was referred as STFT-based SST (FSST) \cite{Hemakom2017}.

Traditional methods like STFT or CWT are restricted by the Heisenberg uncertainty principle, namely, high resolution on both time and frequency domains cannot be achieved simultaneously. Thus, classical linear TF analysis methods generate a “blurred” TF representation, failing to characterize TF features of non-stationary signals accurately \cite{LI2020107243}. SST is related to the class of time-frequency reassignment (TFR) post-processing algorithms that are used for IF estimation from the modulus of a TF representation. TFR methods apply a reassignment map that concentrates and sharpens the spectrogram energy around the IF curves, resulting in a pointed TF plot \cite{THAKUR20131079}. Unlike classical TFR techniques, SST allows for the reconstruction of the components \cite{Hemakom2017}. Hence, SST has been successfully applied to analyze non-stationary signals in several applications, such as medical electrocardiography (ECG) reading \cite{WU2014354}, atomic crystal images in physics \cite{Yang2015_1,LU2016194}, mechanical engineering \cite{Tu2019,HU2019126,Wang2017}, art investigation \cite{Yang2015_2,Cornelis2017}, geology \cite{Yang2014}, etc. Previously, SST has been applied in the field of power systems for parameter identification of low frequency natural oscillations \cite{Wang2018}, and subsynchronous oscillation detection \cite{He2015}. In both cases, the method was applied to simulated data of reduced power system models. Nevertheless, the SST has not been applied yet to analyze FOs in electrical power systems using PMU data.

In this paper, we propose a novel methodology by using the FSST to extract non-stationary components of power system FOs. This technique is applied to simulated as well as real PMU data. Then we apply the DEF Method to the extracted components to trace the source of the FO. The purpose of the proposed methodology is to provide a systematic framework for the application of the DEF method in the case of non-stationary FOs.

The rest of this paper is organized as follows. Background on DEF Method and SST is presented in Section II. Section III presents the proposed methodology. In Section IV the methodology is validated by applying it to measured and simulated power system data, and also it is compared with the usual application of the DEF Method. Concluding remarks are outlined in Section V.

\section{Background on Dissipation Energy Flow Method and Synchrosqueezing Transform}

\subsection{Conventional Application of The Dissipation Energy Flow Method}
The first steps of the DEF method are \cite{MASLENNIKOV201755}:

\begin{itemize}
\item \textit{PMU data pre-processing}. PMU angles for voltages and currents should be unwrapped. Replace missing PMU data (NaN) and outliers with interpolated data. Do unwrapping before interpolation.
\item \textit{Frequency identification}. Use the Discrete Fourier Transformation (DFT) to identify $f_s$, the frequency of interest of the sustained oscillation.
\item \textit{Filtering the component of interest}. Band-pass filtering around the identified frequency is applied to the variables of interest for DEF calculation (i.e. active and reactive power, voltage magnitude, angle, and frequency). The filtering process must satisfy the following conditions: (a) preserves phases among all quantities, (b) preserves magnitudes for all filtered quantities.
\end{itemize}

Then, the flow of dissipating transient energy is calculated in the branches of the systems for the filtered components. The rate of change of the dissipating energy flow allows tracing the source of sustained oscillations. The flow of dissipating energy, for specific filtered components in a branch $ij$, is expressed by integrating over the system trajectory as follows \cite{MASLENNIKOV201755}:

\begin{equation}
W_{ij} \approx \int \Delta P_{ij} \, d\Delta \theta_i + \int \Delta Q_{ij} \frac{d\Delta V_i}{V_i},
\end{equation}
where $P_{ij}$ and $Q_{ij}$ are the active and reactive power flows in branch $ij$, $\theta_i$ is the bus voltage angle, and $V_i$ is the bus voltage magnitude. $\Delta$ indicates that the magnitudes are filtered components. For discrete PMU signals, a discrete-time approximation has the form:
\begin{align}
W_{ij,t+1}^D &= W_{ij,t}^D + \Delta P_{ij,t} \, \left(\Delta \theta_{i,t+1} - \Delta  \theta_{i,t} \right) \nonumber \\
&+ \frac{\Delta Q_{ij,t}}{V_{i,t}} \left( \Delta V_{i,t+1} - \Delta V_{i,t} \right) 
\label{eq:flow_disip_energy}
\end{align}
where $t$ represents the time instant. The integration limits are determined from the transient when sustained oscillations have significant magnitude larger than noise.

\subsection{Multicomponent Signal Decomposition with Synchrosqueezing}

We denote by $\hat{s}$ the Fourier transform of function $s$ with the following normalization:
\begin{equation}
\hat{s}(\eta) = \int_{\mathbb{R}} s(x) e^{-j \left[ 2 \pi \eta x
\right]} dx
\label{eq:ft_def}
\end{equation}

The STFT is a local version of the Fourier transform obtained by means of a sliding window $g$ \cite{Oberlin2014}:
\begin{equation}
V_s^g\left(\eta,t \right) = \int_{\mathbb{R}} s(\tau) g(\tau-t) e^{-j \left[ 2 \pi \eta (\tau - t) \right]} d\tau
\label{eq:stft1}
\end{equation}

Non-stationary oscillatory data $s(t)$ is represented by a superposition of oscillatory components as follows \cite{THAKUR20131079}: 
\begin{align}
s(t) &= \sum_{k=1}^{K} s_k(t) +r(t) \label{eq:sum_s}\\
s_k(t) &= A_k(t) \, e^{j2 \pi \, \varphi_k (t)},
\end{align}
where each oscillatory component $s_k(t)$ has a time-varying amplitude $A_k(t)$ and instantaneous frequency (IF) $\varphi'_k(t)$. The signal $r(t)$ is a noise signal or measurement error plus low frequency trend. If we assume slow variations of $A_k(t)$ and on the IF $\varphi'_k(t)$, we can write the following approximation in the vicinity of a fixed time $t_0$:
\begin{align}
s(t) \approx \sum_{k=1}^{K} A_k(t_0) e^{j2 \pi [\varphi_k (t_0)+\varphi'_k(t)(t-t_0)]}
\end{align}

The corresponding approximation for the STFT then results (changing $t_0$ by a generic $t$):
\begin{align}
V_s^g\left(\eta,t \right) \approx \sum_{k=1}^{K} s_k(t) \hat{g}(\eta-\varphi'_k(t))
\end{align}

The representation of this multicomponent signal by $V_s^g\left(\eta,t \right)$ in the TF plane shows that the peaks are concentrated around so-called ridges, defined by $\eta=\varphi'_k(t)$. The frequency width around each ridge is related to the frequency bandwidth of $\hat{g}$. If frequencies $\varphi'_k(t)$ are separated enough for different $k$, each component has a distinct domain in the TF plane, allowing for their detection, separation and reconstruction \cite{Oberlin2014}. The aim of the SST is twofold. On one hand, it provides a concentrated representation of multicomponent signals in the TF plane. On the other hand, it is a decomposition method that enables the separation and demodulation of the different components. Starting from STFT, the FSST moves the coefficients $V_s^g\left(\eta,t \right)$ according to the map  $\left( \eta, t \right) \mapsto \left( \overline{f} \left( \eta, t \right), t \right)$, where $\overline{f} \left( \eta, t \right)$ is defined by \cite{Oberlin2014}:

\begin{align}
\overline{f} \left( \eta, t \right) &= \frac{1}{2 \pi} \partial_t \,
 arg \left\lbrace V_s^g\left(\eta,t \right) \right\rbrace \nonumber \\
 & = \eta + \Im{\left\lbrace \frac{V_s^{g'} \left(\eta,t \right)}
 {2 \pi\, V_s^g\left(\eta,t \right)} \right\rbrace}
\label{eq:stft2}
\end{align}
where $g'$ is the time derivate of the sliding window $g$. The operator defined in (\ref{eq:stft2}) is a local approximation for instantaneous frequency $\varphi'_k(t)$ at time $t$, filtered at frequency $\eta$. Then, the short-time Fourier transform-based synchrosqueezing transform (FSST) coefficients are given by \cite{Oberlin2014}:
\begin{equation}
T_s^g\left(f,t \right) = \frac{1}{g(0)} \int_{0}^{\infty} V_s^g\left(\eta,t \right) \, \delta\left[f - \overline{f} \left( \eta, t \right) \right] \, d\eta 
\label{eq:stft3}
\end{equation}
where $\delta$ denotes the Dirac distribution. The FSST sharpens the information relative to components in the TF plane around the ridges associated to $\varphi'_k(t)$. Each component $s_k$ can be recovered by integrating $T_s^g\left(f,t \right)$ around a small frequency band $d$ around the curve of $\varphi'_k(t)$ associated to the $k$th component \cite{Oberlin2014}:
\begin{equation}
s_k(t) = \int_{|f-\varphi'_k(t)|<d} T_s^g(f,t) \, df
\label{eq:stft4}
\end{equation}

Practical implementation of the SST algorithm is described in \cite{THAKUR20131079}. Theoretical foundation of SST for oscillatory component extraction can be found in \cite{DAUBECHIES2011243,Oberlin2014}. FSST theoretical foundation was established assuming $g$ as a Gaussian window and the following assumptions on the signal $s(t)$ \cite{Auger2013}:

\begin{description}
\item [{A1)}] $s_k$ has weak frequency modulation, implying the existence of small $\epsilon$ such that for each $t$, one has $\sigma^2 |\varphi''_k(t)| \leq \epsilon$ and $|A'_k(t)| \leq \,\epsilon\,\varphi'_k(t)$, where $\sigma$ is the standard deviation of the Gaussian window.
\item [{A2)}] The components are well separated in frequency. If the frequency bandwidth of $g$ (in rad/s) is $\left[ -\Delta, \Delta \right]$ ( $\Delta=\sqrt{2 \log(2)}/\sigma$ since $g$ is the Gaussian window), this assumption corresponds to the inequality $|\varphi'_k(t)-\varphi'_l(t)| \leq \,2 \, \Delta$ for each $t$ and $k \neq l$.
\end{description}

\section{Proposed Methodology}

\begin{figure}[h]
\centering
\vspace*{-2mm}
\includegraphics[width = 0.9\columnwidth]{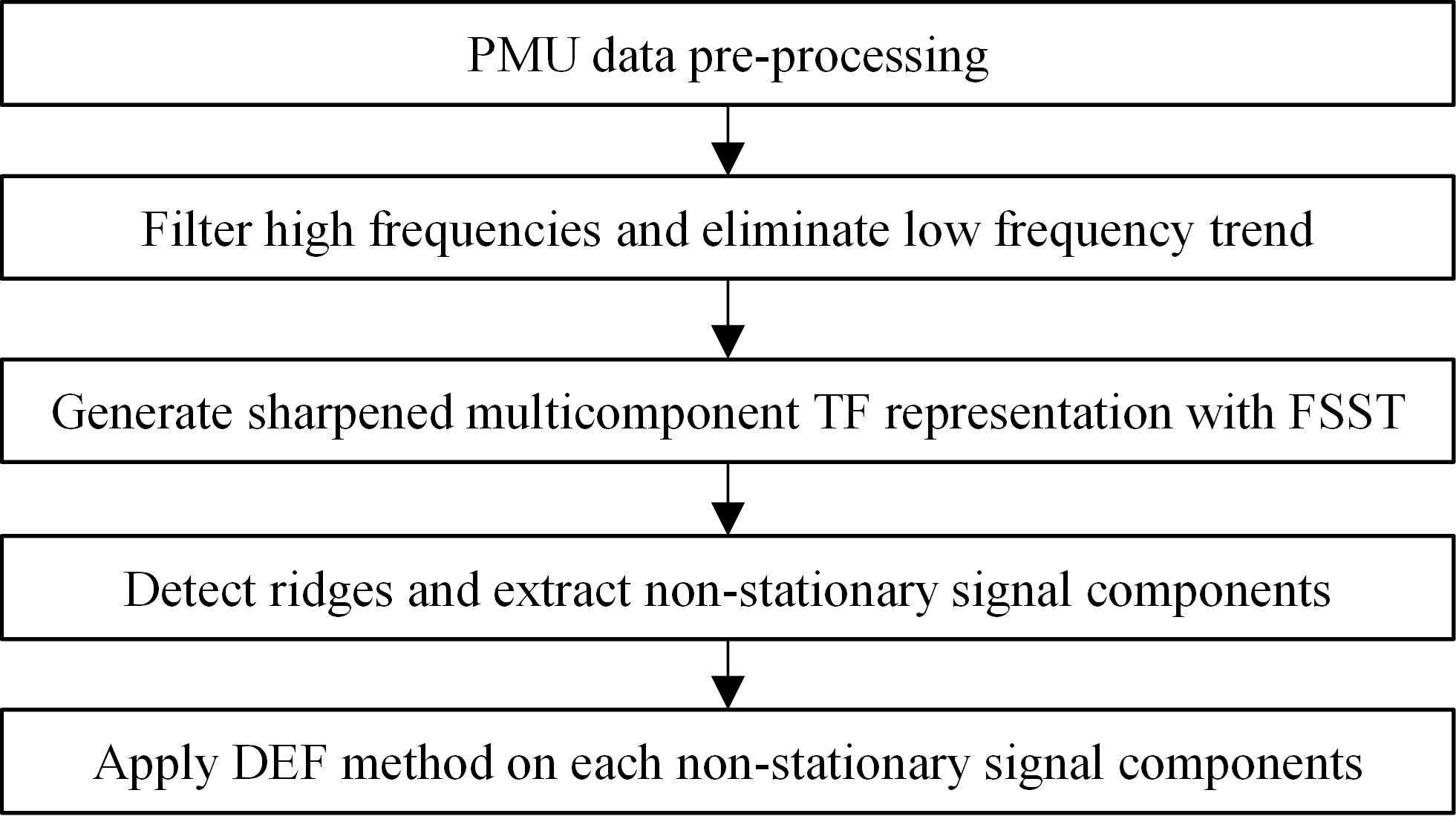}
\vspace*{-2mm}
\caption{Scheme of Proposed Methodology.}
\vspace*{-2mm}
\label{fig:Scheme}
\end{figure}

Fig. \ref{fig:Scheme} shows the scheme of the proposed methodology, which consists of five steps. 
In the same way as in the conventional application of the DEF method, the first step is PMU data pre-processing, where PMU angles are unwrapped and missing PMU data (NaN) and outliers are replaced with interpolated data.
The second step is high frequency filtering and elimination of the low frequency trend. Several methods can be used in this step. In particular, in this paper noise-assisted multivariate empirical mode decomposition (NA-MEMD) \cite{Mandic2013} is used as data-driven band-pass filter to subtract low frequency trend and high frequency content. It is an extension of the EMD [10] to multivariate cases, with the assistance of noise in order to enforce the dyadic filterbank property \cite{Mandic2013}. Code for implementing NA-MEMD is available in \cite{Mandic2017}. Due to the dyadic filter bank property of NA-MEMD, this method was not useful to separate non-stationary components, and mode mixing occurs. However, NA-MEMD is useful as an initial band-pass filter over the range of frequencies of interest \cite{Hemakom2017}. 
In the third step, FSST is applied using (\ref{eq:stft1}), (\ref{eq:stft2}) and (\ref{eq:stft3}) to each signal to generate multiple univariate multicomponent TF planes with high localization in both time and frequency. We use a Guassian window and we seek the value of the standard deviation $\sigma$ leading to the most concentrated TF representation. Spectrum concentration could be measured for example by means of Shannon entropy, Rényi entropy or by the "energy on the ridge" approach \cite{Meignen2017}. In this paper, standard deviation of the window is chosen to minimize the third order Rényi entropy, defined as \cite{Baraniuk2001}
\begin{equation}
H_s^g =-\frac{1}{2} \log_2 \frac{\iint |T_s^g(f,t)|^3 \, df \, dt}{\iint |T_s^g(f,t)| \, df \, dt}
\label{eq:renyi}
\end{equation}
In the fourth place, ridge identification and signal component reconstruction is performed using (\ref{eq:stft4}). A standard ridge identification method is used, where a penalized forward-backward greedy algorithm is used to extract the maximum-energy ridges from a time-frequency matrix. The algorithm finds the maximum time-frequency ridge by minimizing $-\ln|T_s^g(\eta,t)|$ at each time point. Minimizing $-\ln|T_s^g(\eta,t)|$ is equivalent to maximizing the value of $|T_s^g(\eta,t)|$. The algorithm optionally constrains jumps in frequency with a penalty that is proportional to the distance between frequency bins \cite{Meignen2016, MathWorks1}. The description of the functions used for FSST can be found in \cite{MathWorks2} and for ridge extraction in \cite{MathWorks1}. 
Finally, DEF Method is applied using (\ref{eq:flow_disip_energy}) locally in each substation over each non-stationary oscillating component for oscillation source tracing.

\section{FO Analysis}

In this section, we present the analysis of two different examples. First, the proposed methodology is applied to simulated data obtained from the WECC179 model. Secondly, the proposed methodology is applied to real PMU measurements obtained from an event occurred in ISO-NE \cite{Maslennikov2018}. This case shows that non-stationary FOs may exist in real power system operation and it motivates the development of the methodology presented here. In both cases, the performance of the proposed methodology is compared with the application of the DEF method based on the identification of frequencies by DFT and subsequent extraction of the signal components through the design of band pass filters, according to \cite{MASLENNIKOV201755}.

\subsection{Simulated Case.}

\begin{figure}[H]
\centering
\vspace*{-2mm}
\includegraphics[width= 0.75\columnwidth]{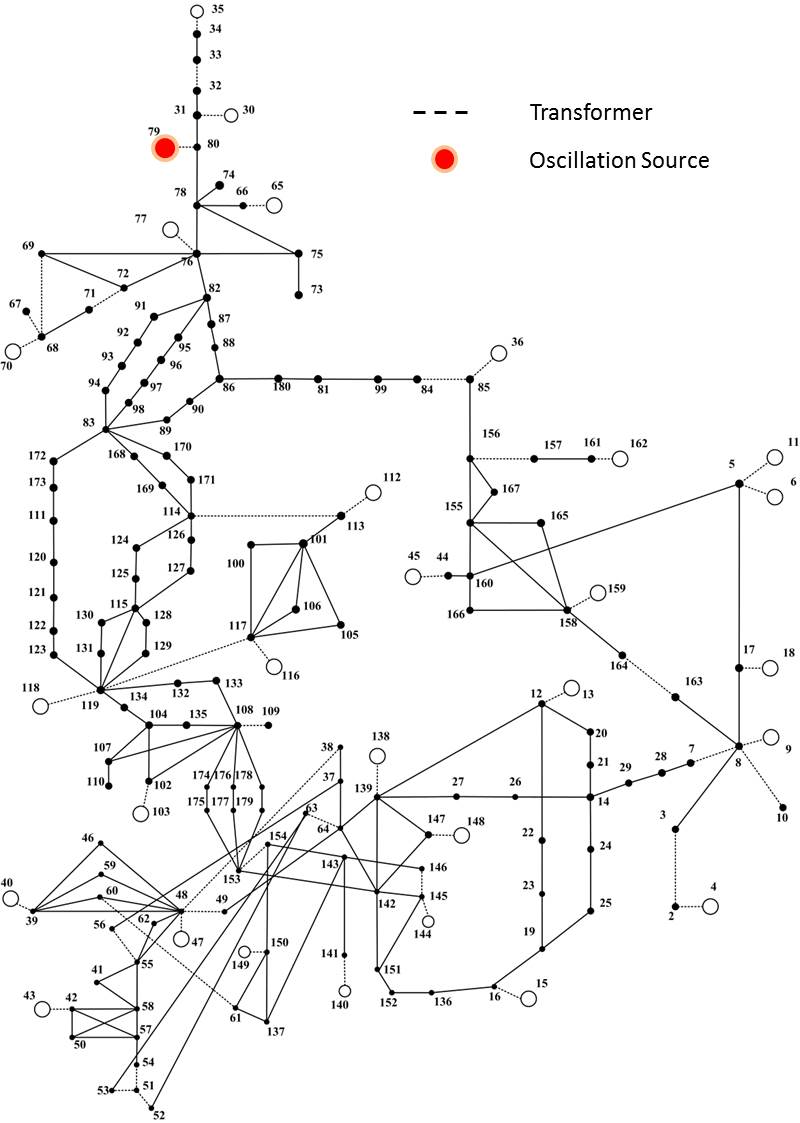}
\vspace*{-2mm}
\caption{WECC179 Test System \cite{Maslennikov2018}.}
\vspace*{-2mm}
\label{fig:WECC_179}
\end{figure}

Fig. \ref{fig:WECC_179} shows the WECC 179 test system. Data for simulation were obtained from  \cite{Maslennikov2018}. 

\begin{figure}[h]
\centering
\vspace*{-5mm}
\includegraphics[width=1\columnwidth]{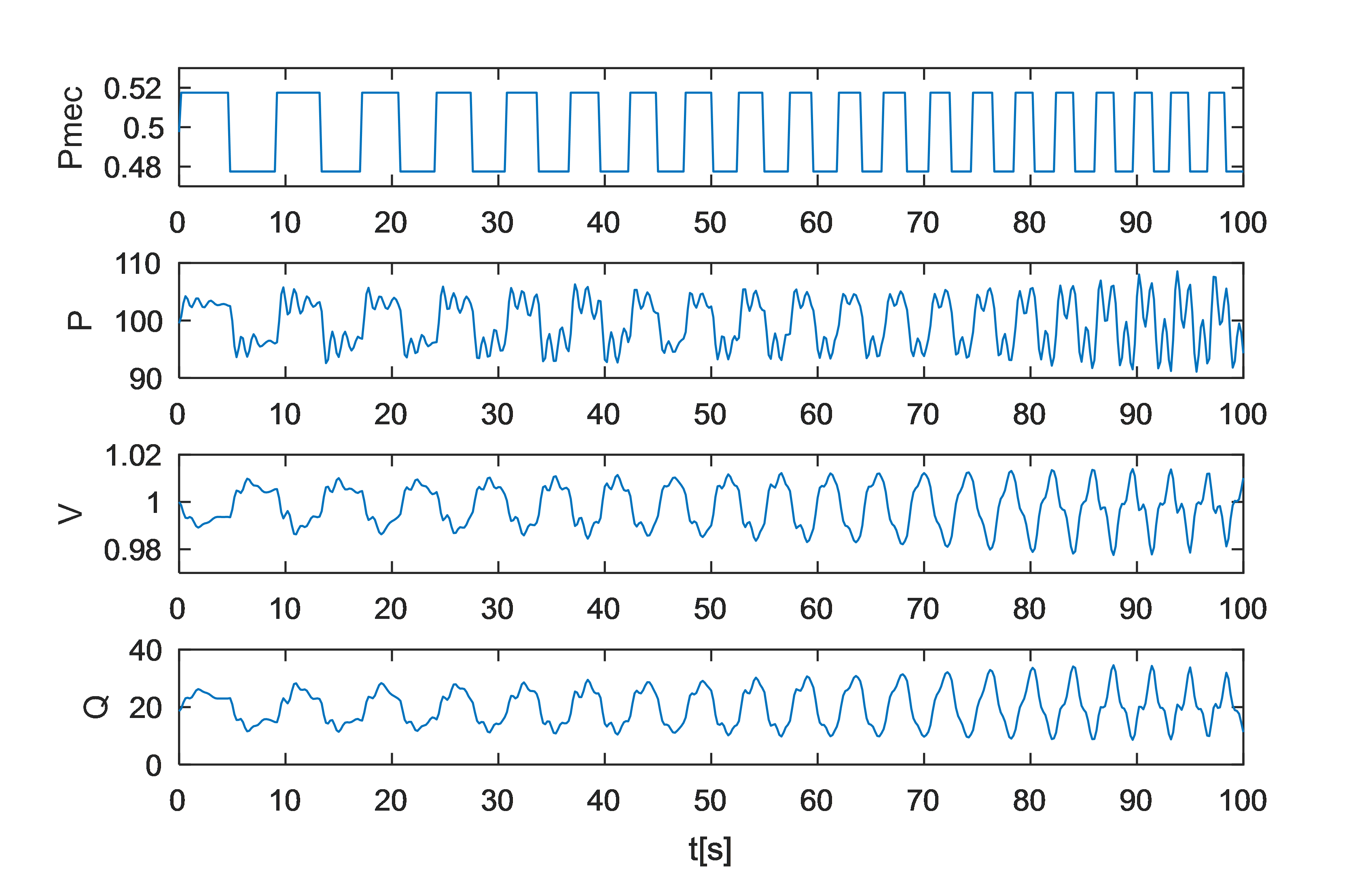}
\vspace*{-5mm}
\caption{Generator 79 mechanical power Pmec in pu of generator base (20000 MVA), electric power P in pu of system base (100 MVA), voltage magnitude V in pu and reactive power Q in pu of system base 100 MVA.}
\vspace*{-2mm}
\label{fig:Gen_signals}
\end{figure}

A non-stationary mechanical power was applied in generator 79 (indicated as the oscillation source in Fig. \ref{fig:WECC_179}) using a square signal whose fundamental frequency is linearly increased from 0.1 Hz to 0.3 Hz in 100s. Fig. \ref{fig:Gen_signals} shows the mechanical power $P_{mec}$, electric power $P$, terminal voltage magnitude $V$, reactive power $Q$ of generator 79.

\subsubsection{Simulated Case. Proposed Methodology based on FSST}

\begin{figure}[h]
\centering
\vspace*{-2mm}
\includegraphics[width=1\columnwidth]{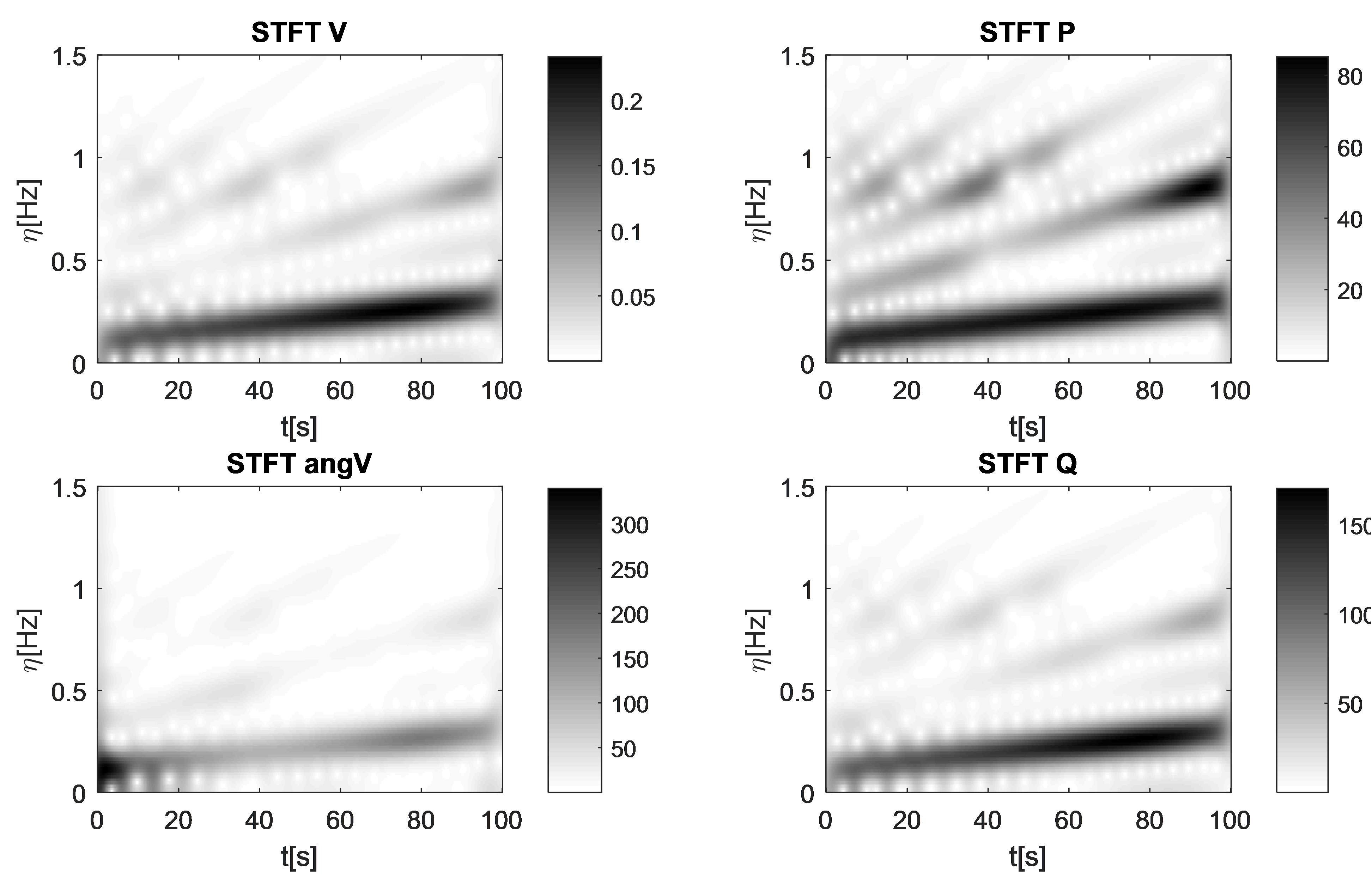}
\vspace*{-5mm}
\caption{STFT of magnitudes of generator 79 for $\sigma=2.5$s.}
\vspace*{-2mm}
\label{fig:Specs_STFT}
\end{figure}

\begin{figure}[h]
\centering
\vspace*{-2mm}
\includegraphics[width=1\columnwidth]{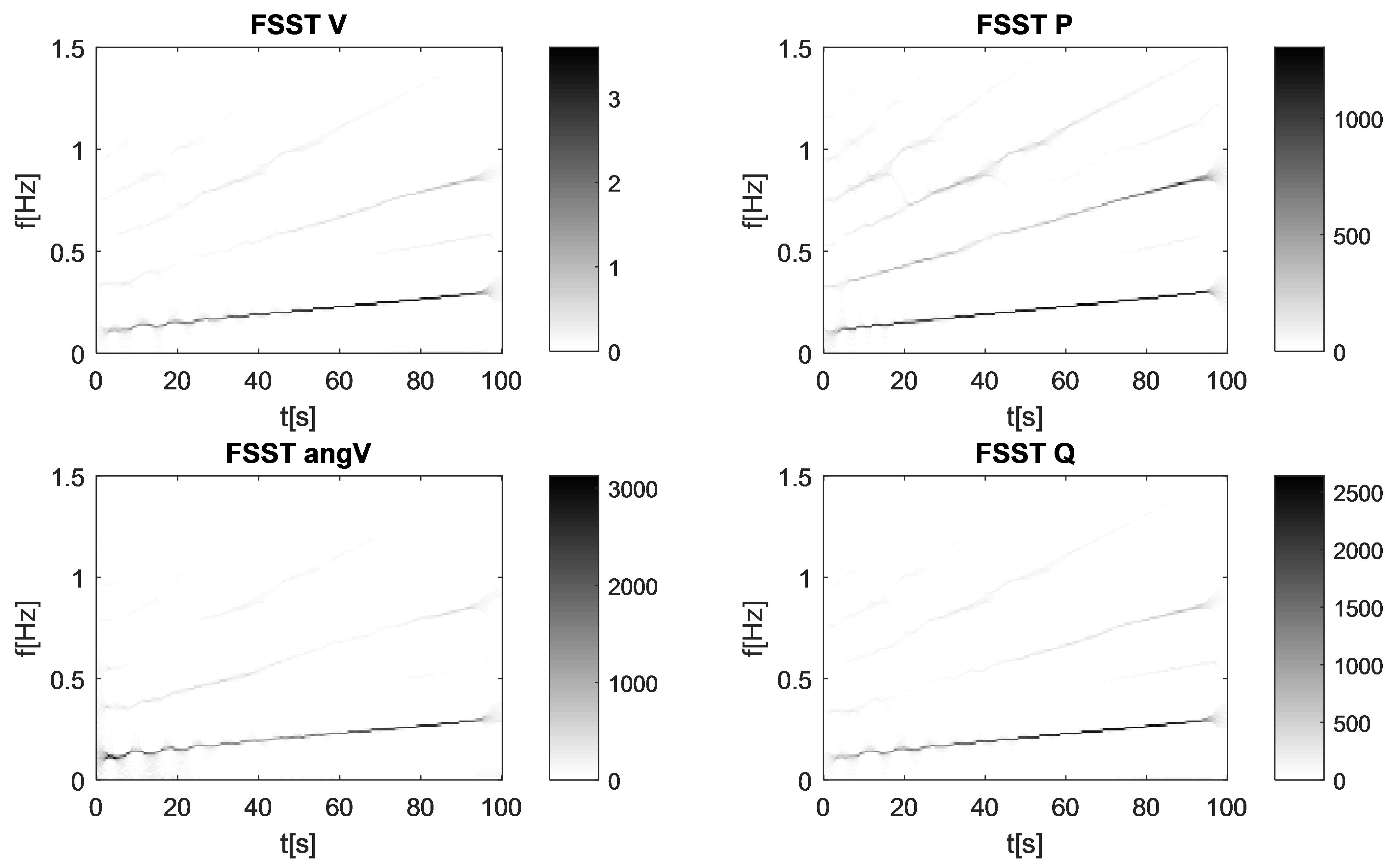}
\vspace*{-5mm}
\caption{FSST of magnitudes of generator 79 for $\sigma=2.5$s.}
\vspace*{-2mm}
\label{fig:Specs_FSST}
\end{figure}

\begin{figure}[H]
\centering
\vspace*{-2mm}
\includegraphics[width=1\columnwidth]{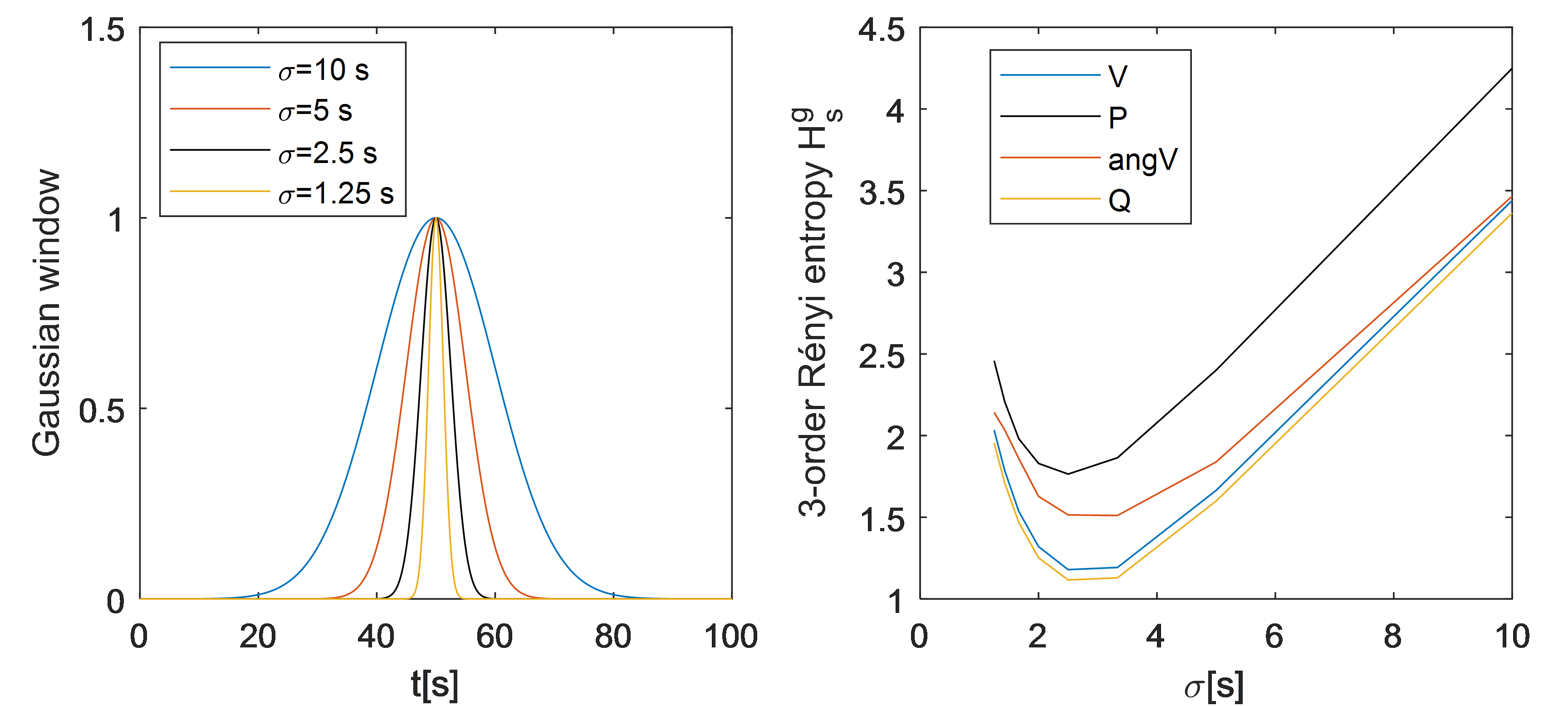}
\vspace*{-5mm}
\caption{Gaussian window (left) and Rényi entropy of magnitudes of generator 79 (right) for different values of $\sigma$.}
\vspace*{-2mm}
\label{fig:ventana}
\end{figure}

\begin{figure}[H]
\centering
\vspace*{-2mm}
\includegraphics[width=1\columnwidth]{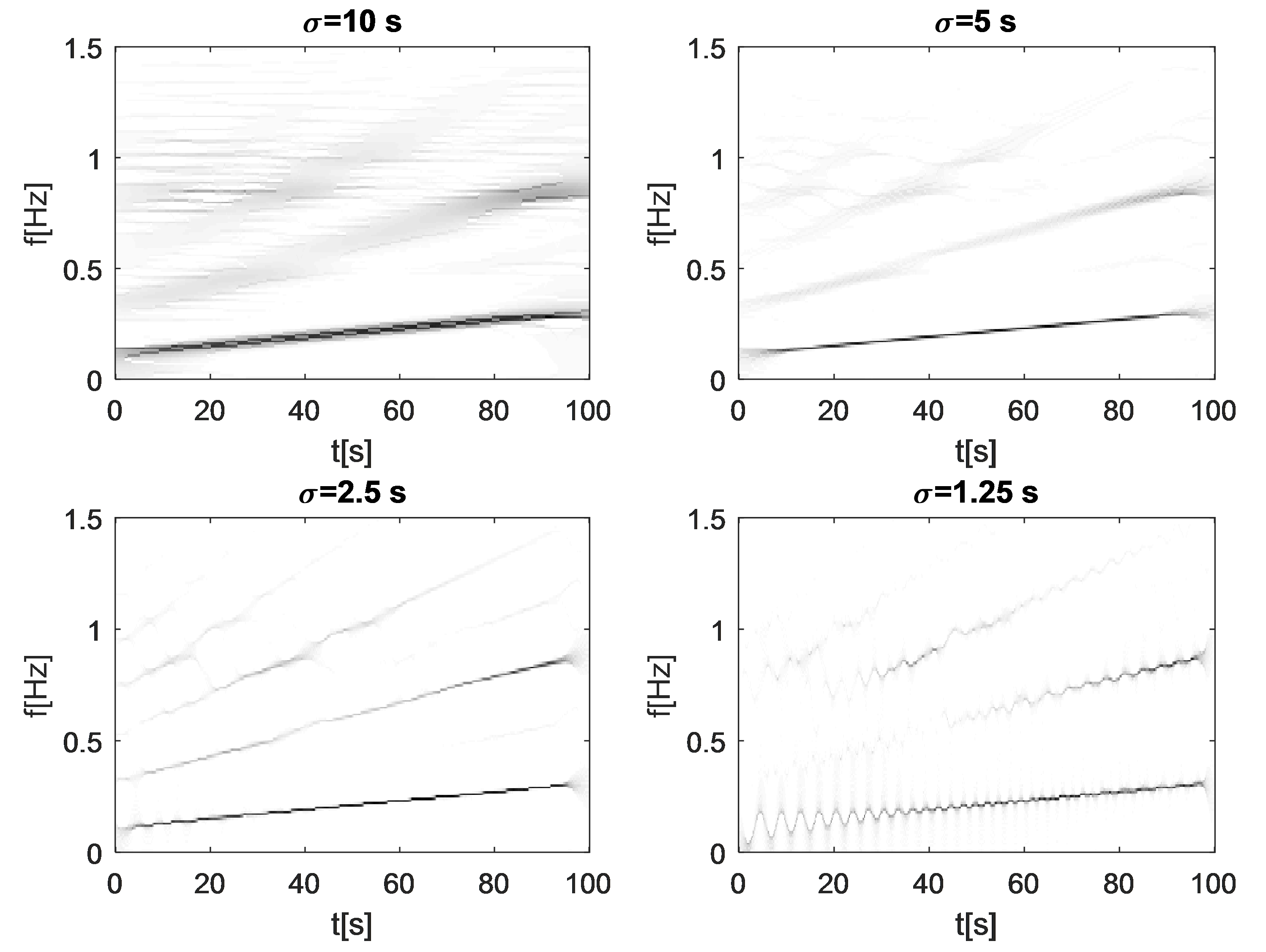}
\vspace*{-5mm}
\caption{FSST of $P$ of generator 79 for different values of $\sigma$.}
\vspace*{-2mm}
\label{fig:Espectro_sigma}
\end{figure}

In this simulated case, the elimination of low frequency trend consists in extracting the mean value of the signals. For the third step of proposed methodology, STFT is first calculated with (\ref{eq:stft1}). Fig. \ref{fig:Specs_STFT} shows amplitude of STFT over TF plane for $P$, $V$, $Q$ and terminal voltage angle $angV$ of generator 79, using Gaussian window with $\sigma=2.5$s. It can be seen that STFT concentrates the information around the ridges that correspond to the time-varying fundamental frequency and harmonic frequencies of the forced oscillation. Then, FSST is calculated using (\ref{eq:stft3}). Fig. \ref{fig:Specs_FSST} shows that FSST improves the definition around each ridge allowing better identification of the components in the time-frequency plane. Window standard deviation of $\sigma=2.5$s was selected considering minimization of Rényi entropy calculated with (\ref{eq:renyi}), as shown in Fig. \ref{fig:ventana}. FSST of electric power $P$ for different values of $\sigma$ is shown in Fig. \ref{fig:Espectro_sigma}, reflecting that the one with the least entropy is also the one that best defines the ridge curves.

\begin{figure}[h]
\centering
\vspace*{-2mm}
\includegraphics[width=0.65\columnwidth]{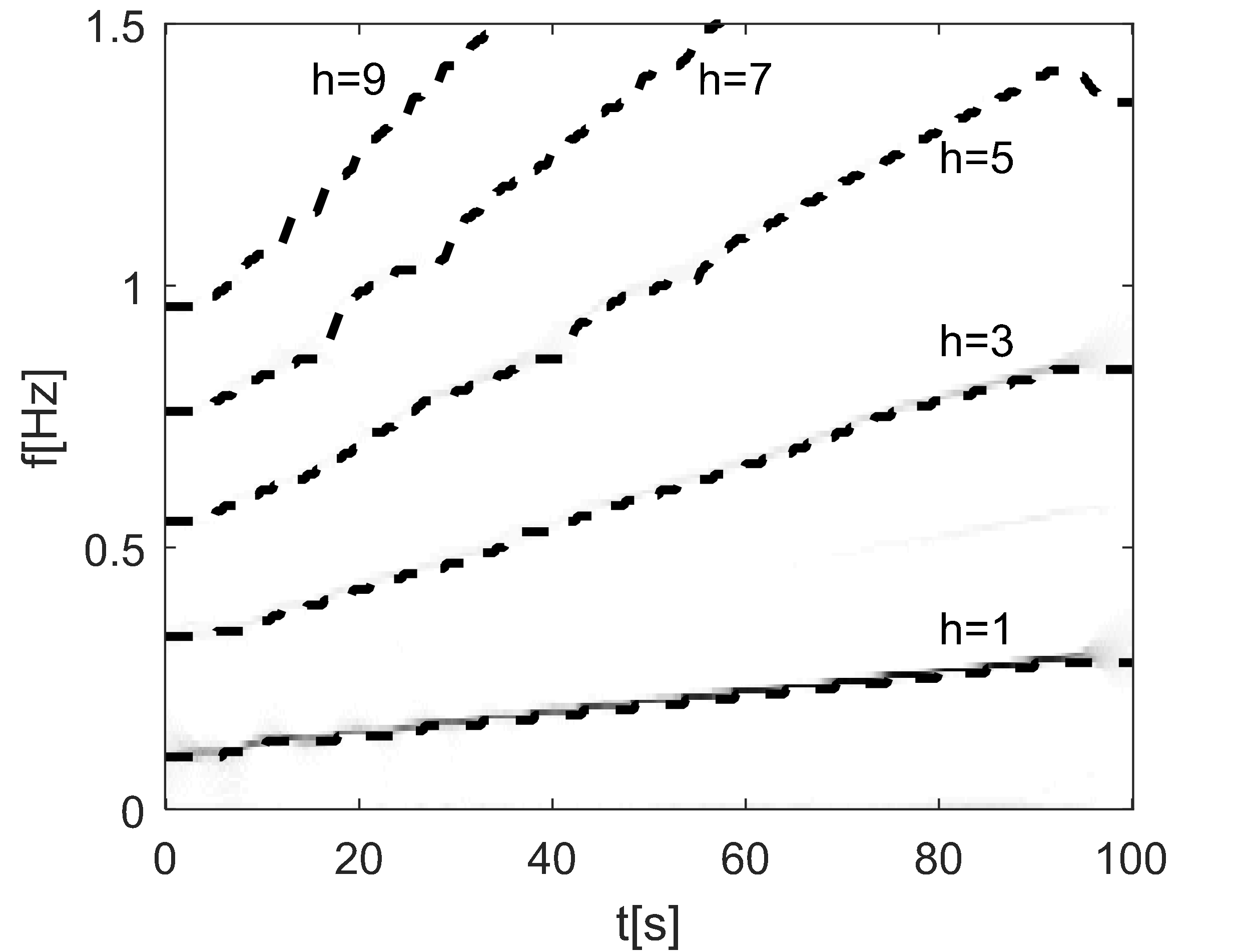}
\vspace*{-2mm}
\caption{Ridge extraction of non-stationary harmonics in generator 79.}
\vspace*{-2mm}
\label{fig:Specs_FSST2}
\end{figure}

\begin{figure}[h]
\centering
\vspace*{-2mm}
\includegraphics[width= 0.97\columnwidth]{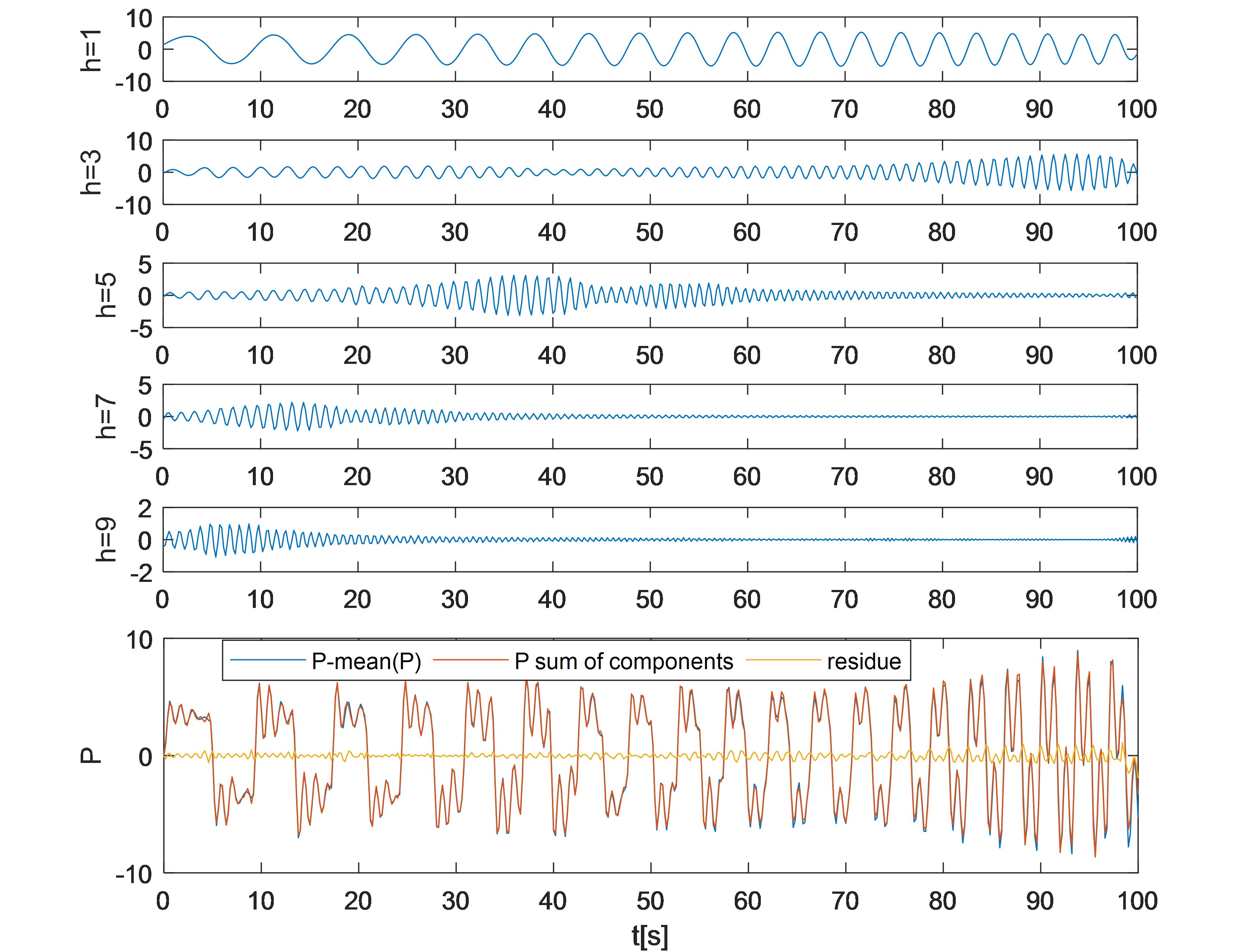}
\vspace*{-3mm}
\caption{Decomposition of electric power P of generator 79}
\vspace*{-2mm}
\label{fig:Descop_modos_simu}
\end{figure}

Fig. \ref{fig:Specs_FSST2}  shows the results of the ridge identification algorithm applied to the FSST amplitude of the active power flow of generator 79. The obtained curves are used for decomposing the other three variables ($V$, $Q$, and $angV$) of the respective generator. For example, Fig. \ref{fig:Descop_modos_simu} shows the extracted $h$-order harmonics applying the reconstruction formula (\ref{eq:stft4}) around each ridge, for the electric power of generator 79. The same procedure is performed on all the 29 generators of the WECC179 test system.

\begin{figure}[H]
\centering
\vspace*{-3mm}
\includegraphics[width= 0.97\columnwidth]{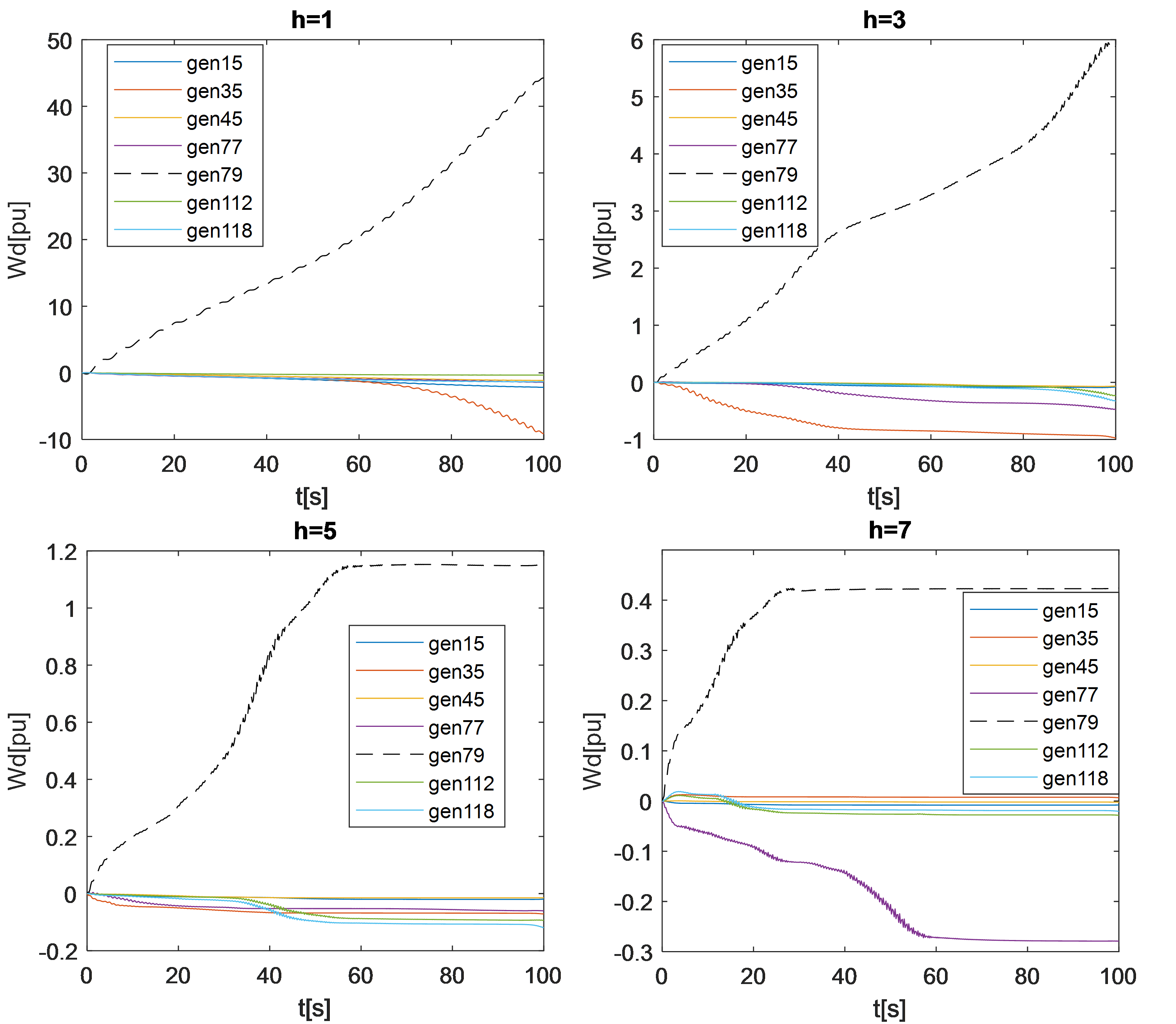}
\vspace*{-3mm}
\caption{DEF of harmonics with higher energy in selected generators.}
\vspace*{-3mm}
\label{fig:DEF_modos_simu}
\end{figure}

After performing the decomposition for all the variables of the generators, we calculate the associated DEF for all harmonics and generators. Fig. \ref{fig:DEF_modos_simu} shows the DEF of harmonics with higher energy for the generators with greater participation in the oscillation process. Looking at the different harmonics, we see that the rate of change of DEF is positive only in generator 79. Thus, it is concluded that generator 79 is the source of oscillations. In particular, harmonics $h=1$ (fundamental) and $h=3$ present significant monotonously increasing DEF values. On the other hand, harmonics $h=5$ and $h=7$ only show DEF increments during certain instants, where the amplitude of the corresponding components is sufficiently large (see Fig. \ref{fig:Descop_modos_simu} and Fig. \ref{fig:DEF_modos_simu} together).

\subsubsection{Simulated Case. DFT and Fixed Band Pass Filtering}

\begin{figure}[h]
\centering
\vspace*{-3mm}
\includegraphics[width= 1\columnwidth]{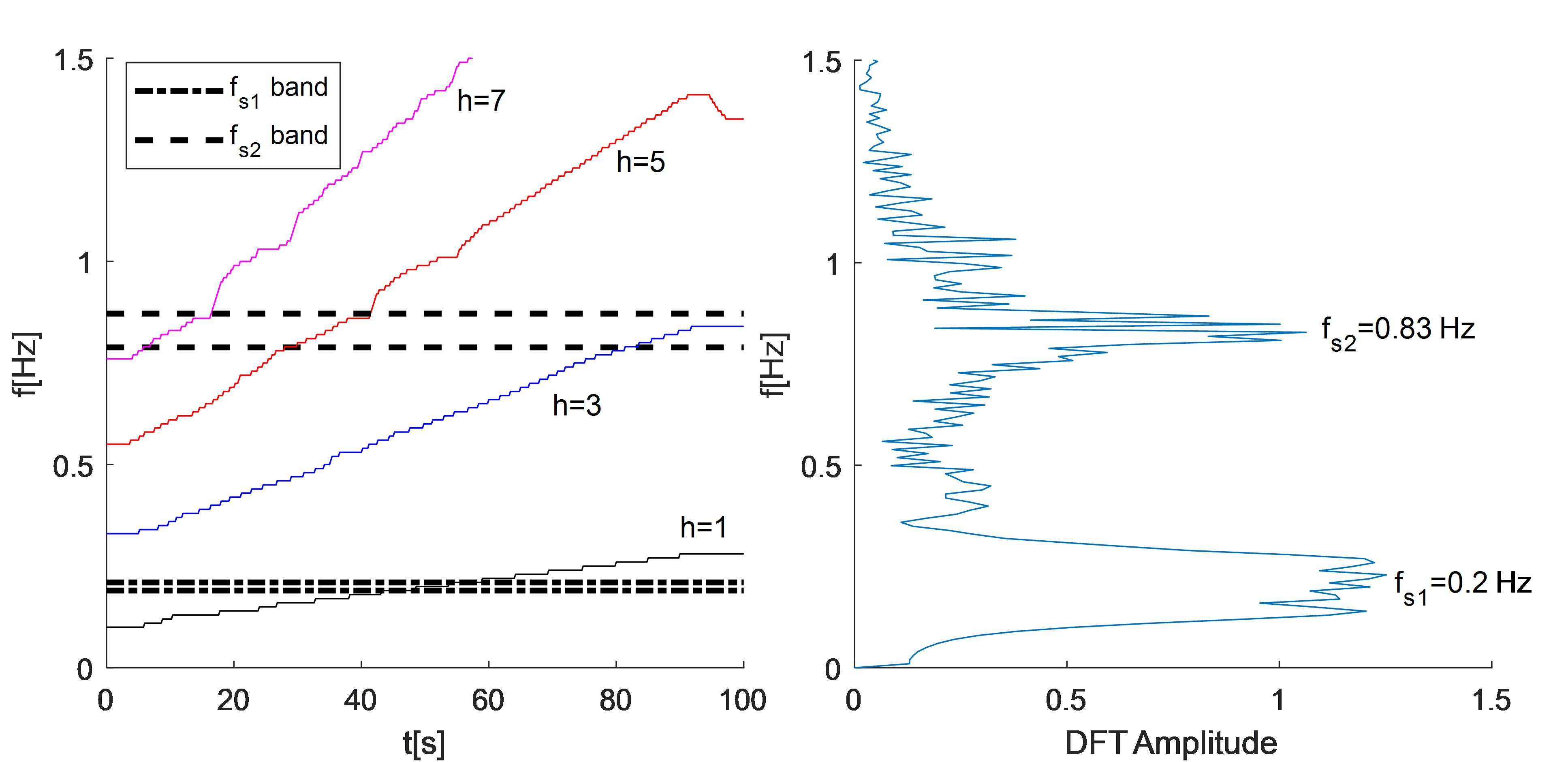}
\vspace*{-3mm}
\caption{DFT of electric power of generator 79.}
\vspace*{-2mm}
\label{fig:FFT_EP}
\end{figure}

\begin{figure}[h]
\centering
\vspace*{-4mm}
\includegraphics[width= 1\columnwidth]{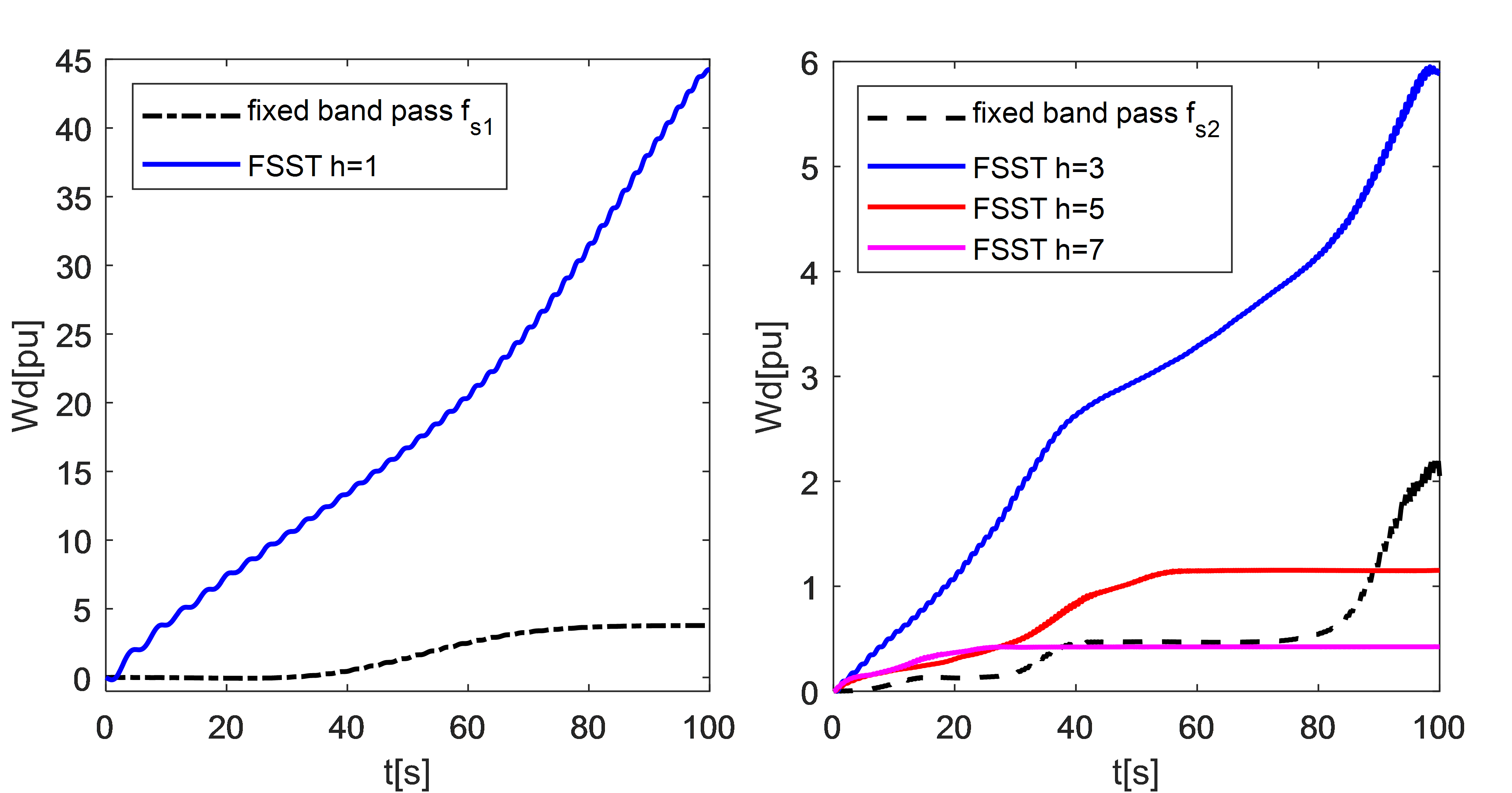}
\vspace*{-3mm}
\caption{Comparison of DEF on Generator 79 obtained with fixed band pass filtering and proposed methodology based on FSST}
\vspace*{-3mm}
\label{fig:DEF_comp}
\end{figure}

The conventional application of the DEF method is intended for signals with stationary behavior in frequency. This section studies that direct application to non-stationary frequency signals could lead to inadequate results. In the first place, DFT of the full time series of the electric active power of generator 79 is calculated, as shown in the right of Fig. \ref{fig:FFT_EP}.  The fundamental frequency cannot be identified as a sharp peak due to the non-stationary behavior. For the application of fixed band-pass filtering, we assume that this wide peak is classified as a frequency component of FO of $f_{s1}$=0.2Hz. On the other hand, a second peak in $f_{s2}$=0.83Hz is identified. Filter design specifications are considered based on \cite{MASLENNIKOV201755}: Butterworth filter with the pass frequencies $f_{pi}=(1 \pm e)f_{si}$ where $e$=0.05; cutoff frequencies $f_{ci}=(1 \pm 2e)f_{si}$; 1 dB of ripple allowed and 10–15 dB attenuation at both sides of the passband. Matlab function \texttt{designfilt} is used for each of the frequencies of interest $f_{si}$ , $i=1,2$. Zero-phase distortion is achieved by applying filtering in both forward and reverse directions for all signals by using the Matlab function \texttt{filtfilt} \cite{MASLENNIKOV201755}. The resulting DEF using a bandpass filter with $f_{s1}$=0.2 Hz is shown at the left of Fig. \ref{fig:DEF_comp}. Here, for comparative purposes, the corresponding fundamental component $h=1$ of the FSST-based decomposition is included. It can be observed that the DEF calculated using the fixed bandpass filtering approach has significant variations only between 40 and 60 seconds. This is the time interval where the instantaneous frequency of the non-stationary component overlaps with the frequency band of the filter, as shown in the left of Fig. \ref{fig:FFT_EP}. Thus, it is unable to capture the complete non-stationary signal component.

DEF calculated from band pass filtered components around frequency $f_{s2}$=0.83Hz is also shown at the right of Fig. \ref{fig:DEF_comp}. It only presents significant values when the filtering band intersects with the non-stationary frequency components, as can be inferred together with Fig. \ref{fig:FFT_EP}. For example, in the interval from 80 to 90 seconds, the frequency band intersects with the instantaneous frequency ridge of third harmonic $h=3$. Only in that time interval, the variations of DEF are similar to those of calculated from FSST-based decomposition. In the same way, this happens with fifth ($h=5$) and seventh ($h=7$) harmonics. In the instants where the filter band does not intercept any instantaneous frequency ridges, the DEF calculated with the fixed band pass filter approach does not suffer variations, resulting in inaccurate results.

\subsubsection{Simulated Case. Windowed DFT and Multiple Band Pass Filtering}

\begin{figure}[H]
\centering
\vspace*{-4mm}
\includegraphics[width= 1\columnwidth]{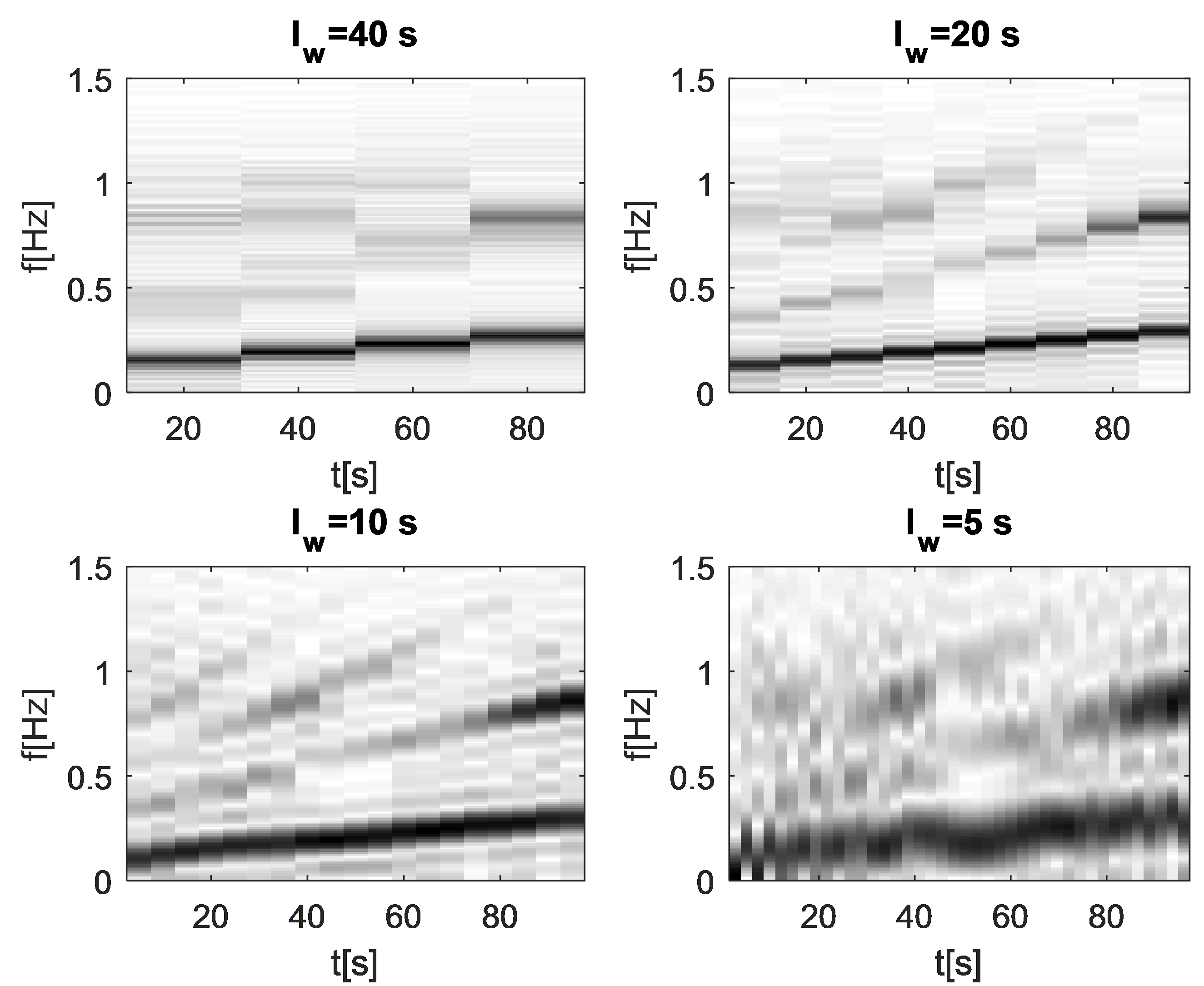}
\vspace*{-5mm}
\caption{DFT of electric power of generator 79 using rectangular sliding window of length l$_w$ with half window overlap}
\vspace*{-3mm}
\label{fig:DFT_lw}
\end{figure}

An extension of the conventional DEF method for analysis of non-stationary signals is considered through the application of windowed DFT for the identification of frequencies and the design of multiple band pass filters, splitting the signal into smaller intervals. Fig. \ref{fig:DFT_lw} shows spectrogram of electric power of generator 79 using DFT with rectangular sliding window of length l$_w$ with half window overlap. It can be seen that the resulting spectrogram with a window of length l$_w$=10 s is very similar to the resulting STFT with Gaussian window with $\sigma$=2.5 s, shown in Fig. \ref{fig:Specs_STFT}.

\begin{figure}[h]
\centering
\vspace*{-4mm}
\includegraphics[width= 0.55\columnwidth]{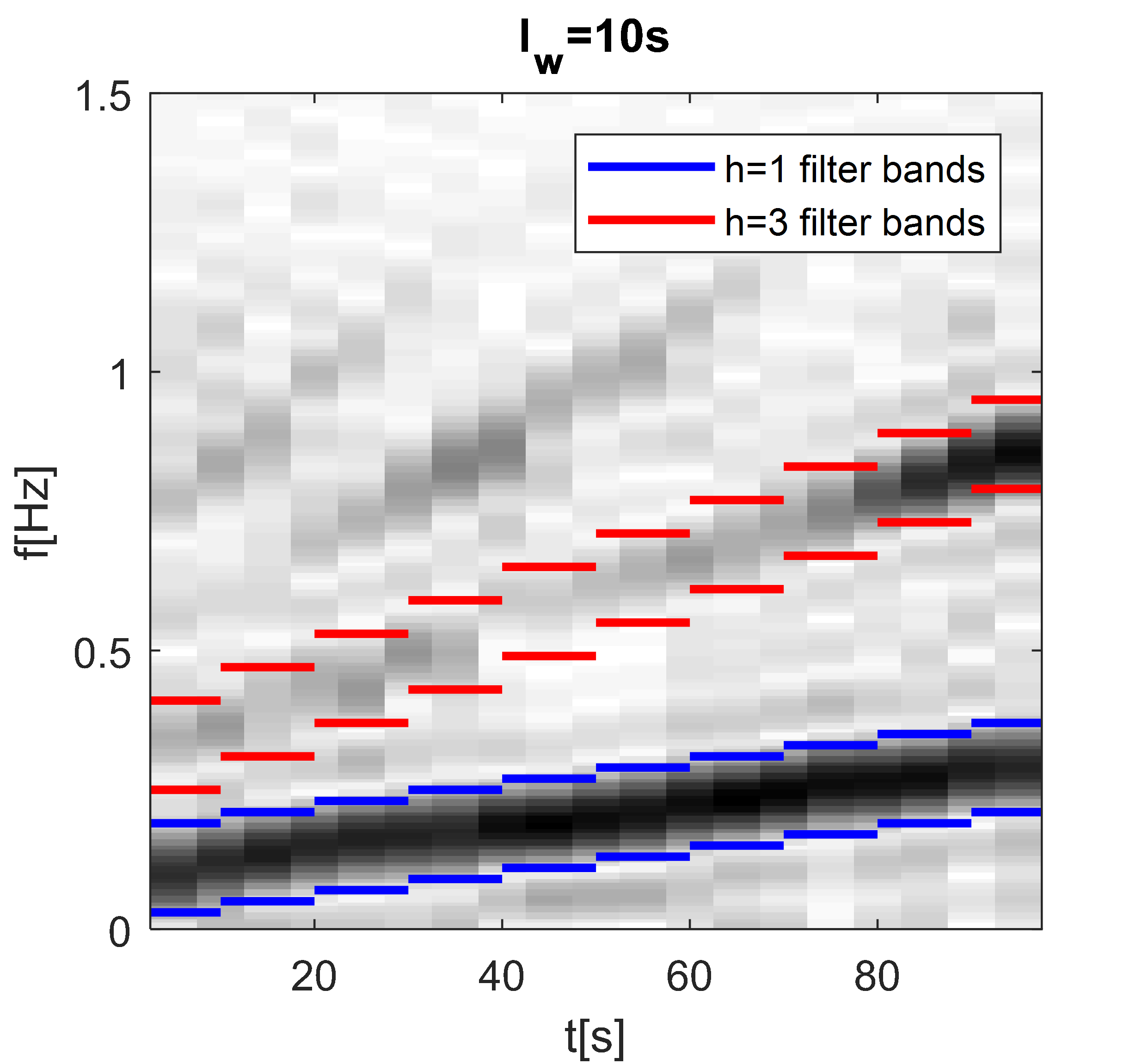}
\vspace*{-3mm}
\caption{Multiple filter bands defined from windowed DFT with rectangular window of length l$_w=10$}
\vspace*{-3mm}
\label{fig:DFT_lw10_conbandas}
\end{figure}

\begin{figure}[h]
\centering
\includegraphics[width= 1\columnwidth]{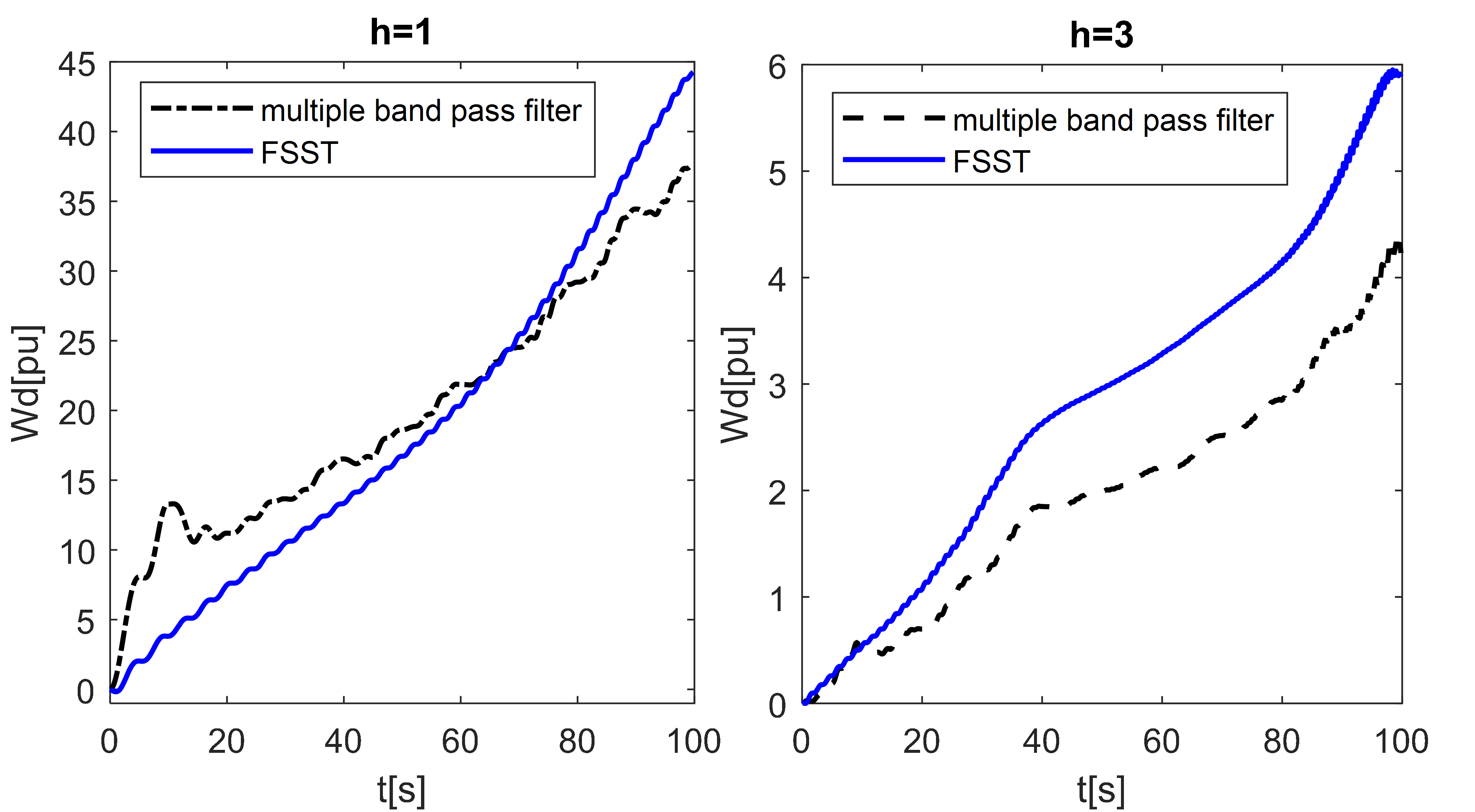}
\vspace*{-5mm}
\caption{Comparison of DEF on Generator 79 obtained with multiple band pass filters and proposed methodology based on FSST}
\vspace*{-1mm}
\label{fig:DEF_comp_var}
\end{figure}

Based on the spectrogram with a square window of l$_w$=10 s, it is proposed to split the signals into 10 second intervals, to which different band pass filters are applied. Fig. \ref{fig:DFT_lw10_conbandas} shows as example the filtering bands for harmonics $h=1$ and $h=3$ adopted from the windowed DFT spectrum. Filters are considered Butterworth type with  1 dB of ripple allowed and 15 dB attenuation at both sides of the passband, as in \cite{MASLENNIKOV201755}.
The resulting DEFs for generator 79, shown in \ref{fig:DEF_comp_var}, are very similar to those obtained with the proposed methodology. 
Although the trend of the results obtained with both approaches is similar, DEF computed from the proposed methodology based on FSTT has a smoother and more monotonous variation.

\subsubsection{Simulated Case. Noise Impact}

\begin{figure}[H]
\centering
\vspace*{-4mm}
\includegraphics[width= 0.97\columnwidth]{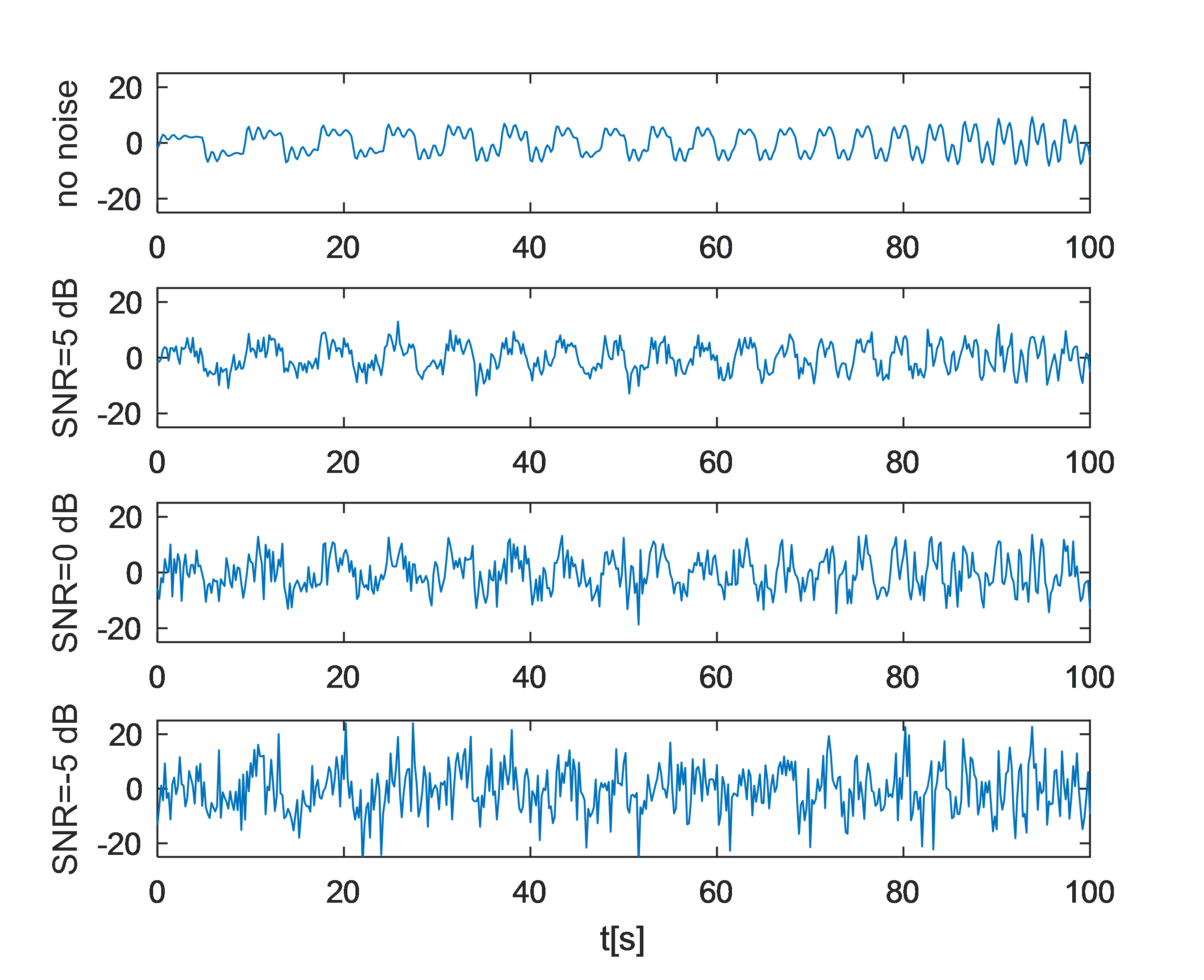}
\vspace*{-3mm}
\caption{Electric power of generator 79 in pu with added noise (the mean value of the signal is subtracted)}
\vspace*{-3mm}
\label{fig:Pelec_SNR}
\end{figure}

Here, we explore the tolerance to noise of FSST, the estimation of ridge curves and the resulting DEF. We denote by $\alpha(t)$ an additive white noise with zero mean and variance $\sigma_\alpha^2$. The Signal-to-Noise Ratio (SNR) measured in dB will be defined by $SNR[dB] =10 \log_{10} (Var(s) / \sigma_\alpha^2)$, where $Var(s)$ is the variance of the noiseless signal $s$ that contains the FO. It should be clarified that this definition of SNR differs from that commonly used to indicate PMU measurement noise, in which the variance of the noise is compared to the signal energy. 
Note that negative SNR values imply that the variance of the noise is greater than that of the FOs components. As example, Fig. \ref{fig:Pelec_SNR} shows electric power of generator 79 for different noise levels. In case of $SNR=0dB$ and $SNR=-5dB$, it results difficult to identify the FO frequencies with the naked eye. 

\begin{figure}[H]
\centering
\vspace*{-3mm}
\includegraphics[width= 1\columnwidth]{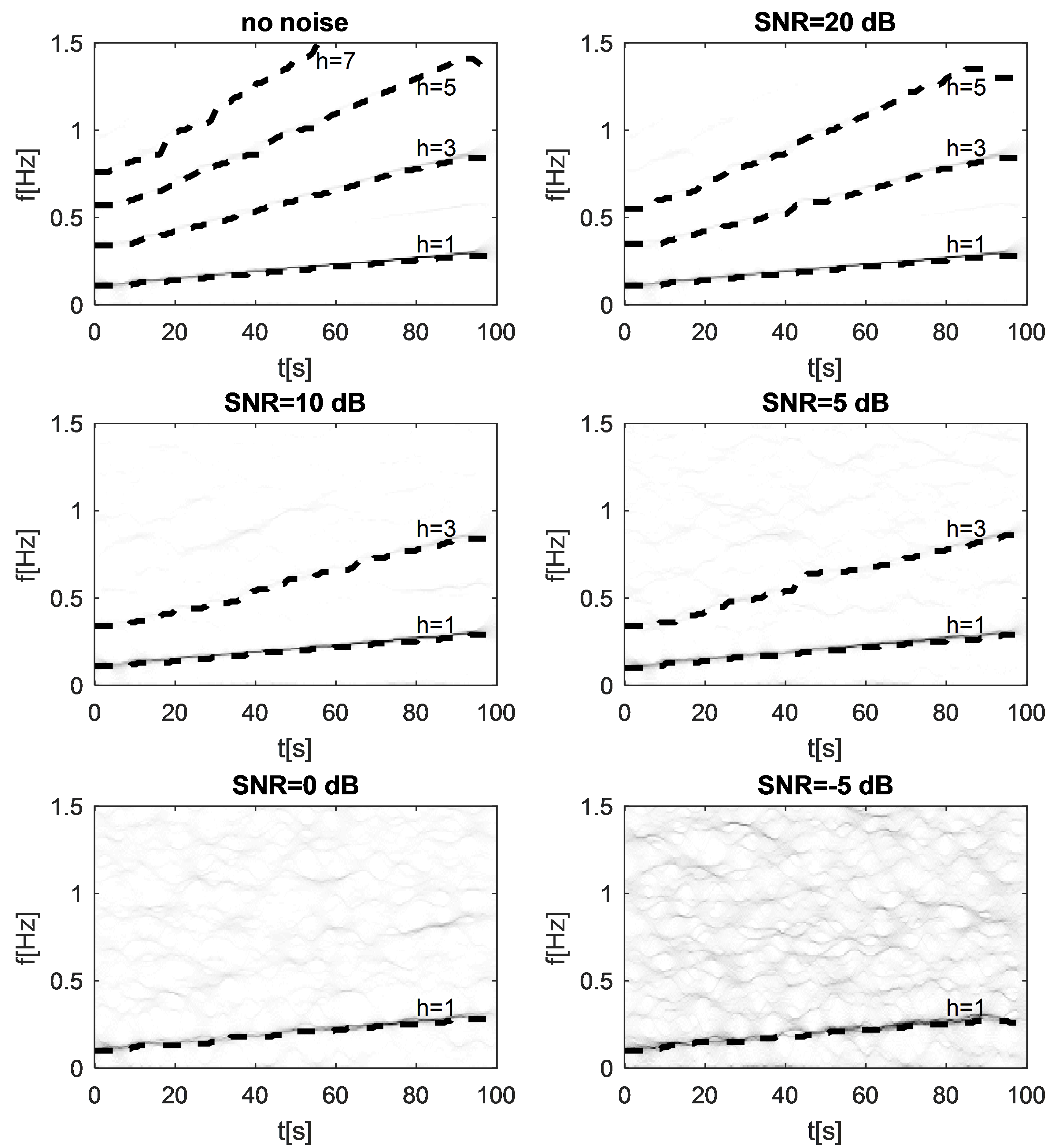}
\vspace*{-5mm}
\caption{Effect of noise on FSST spectrum and identification of ridges}
\vspace*{-3mm}
\label{fig:SST_SNR}
\end{figure}

\begin{figure}[H]
\centering
\vspace*{-4mm}
\includegraphics[width= 0.95\columnwidth]{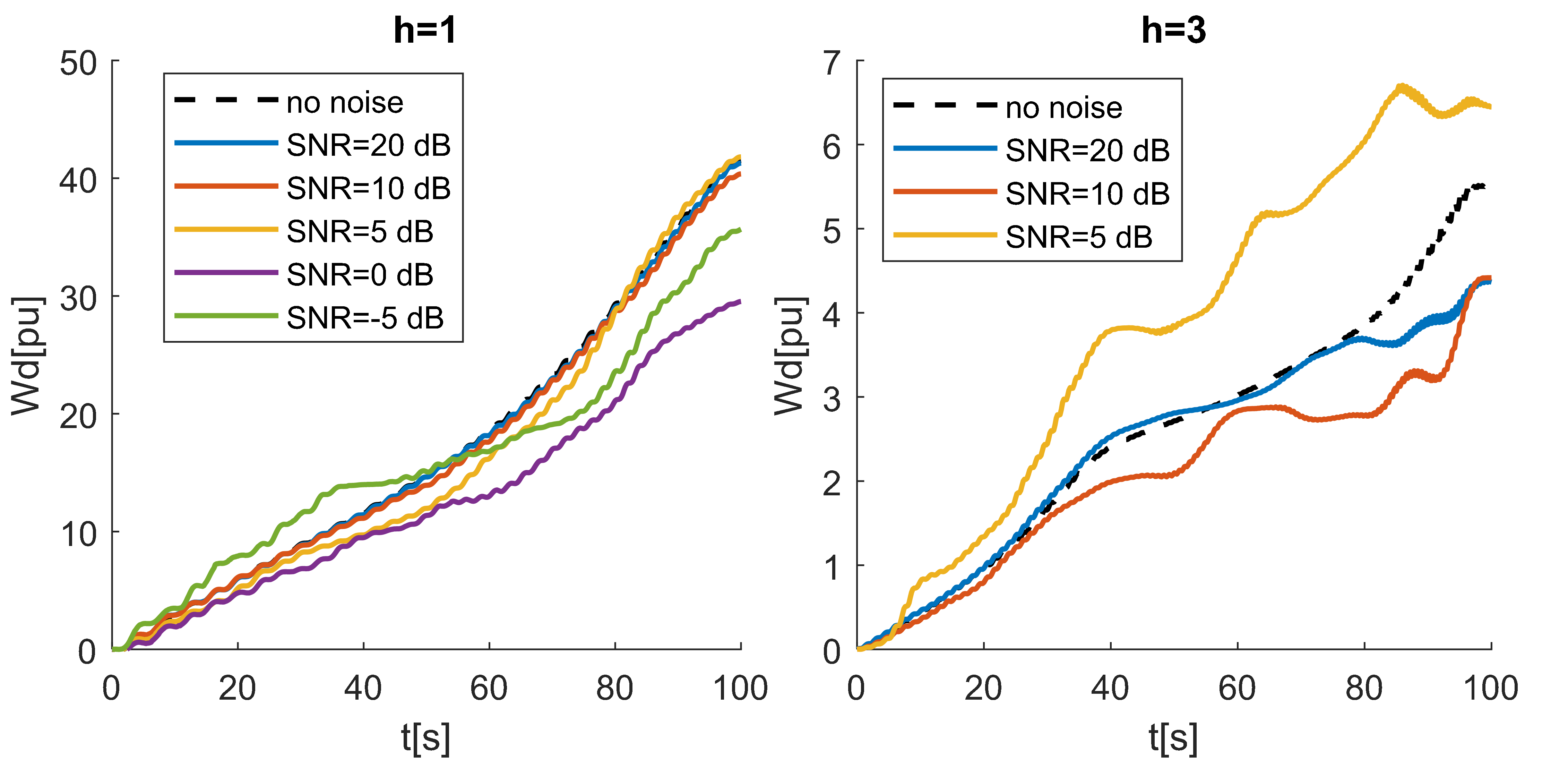}
\vspace*{-3mm}
\caption{DEF for different levels of added noise}
\vspace*{-3mm}
\label{fig:Wd_SNR}
\end{figure}

Fig. \ref{fig:SST_SNR} shows the results of applying FSST algorithm with different levels of added noise to the electric power $P$ of generator 79. It is observed that as the noise intensity increases, the harmonics of lower amplitude cease to be identified. For example, for $SNR = 20dB$ it is not possible to identify $h = 7$, while for $SNR=10dB$ and $SNR=5dB$ only $h=1$ and $h=3$ can be identified. Despite the high noise levels, for $SNR=0dB$ and $SNR=-5dB$, the FSST algorithm can still identify the fundamental component h=1. Fig. \ref{fig:Wd_SNR} shows the resulting DEF with the addition of noise with the indicated level of SNR to each of the magnitudes of the generator 79 ($P$, $Q$, $V$ and $angV$). It is observed that the DEF for the identified harmonics, although it differs slightly from the case without noise, allows to identify generator 79 as a FO source.

\subsection{Event ISO New England System}
\begin{figure}[H]
\centering
\vspace*{-2mm}
\includegraphics[width = 0.9\columnwidth]{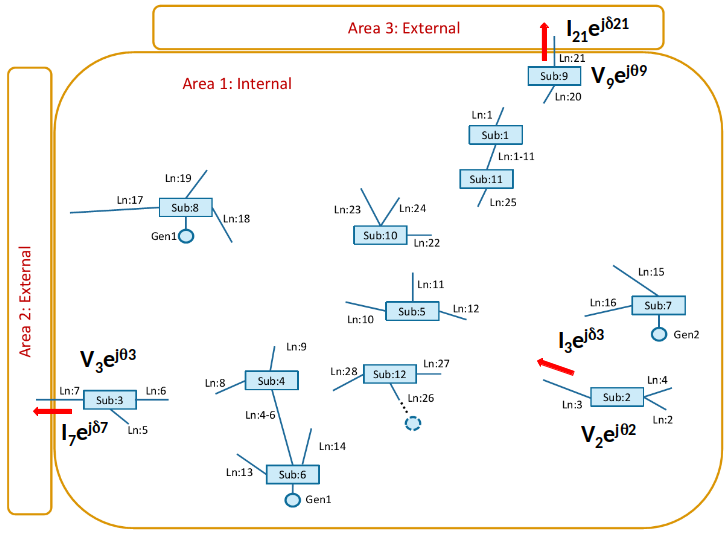}
\vspace*{-2mm}
\caption{ISO-NE Network Scheme and PMU data available [8].}
\vspace*{-2mm}
\label{fig:ISO_NE}
\end{figure}

ISO New England is a North-East part of the Eastern Interconnection in the USA (Peak load is about 26,000 MW). PMU data from real events is available in the Test Cases Library of the IEEE PES Task Force on Oscillation Source Location \cite{Maslennikov2018}. Fig. \ref{fig:ISO_NE} shows the available PMU data and their links to external areas. 

\begin{figure}[h]
\centering
\vspace*{-4mm}
\includegraphics[width = 1\columnwidth]{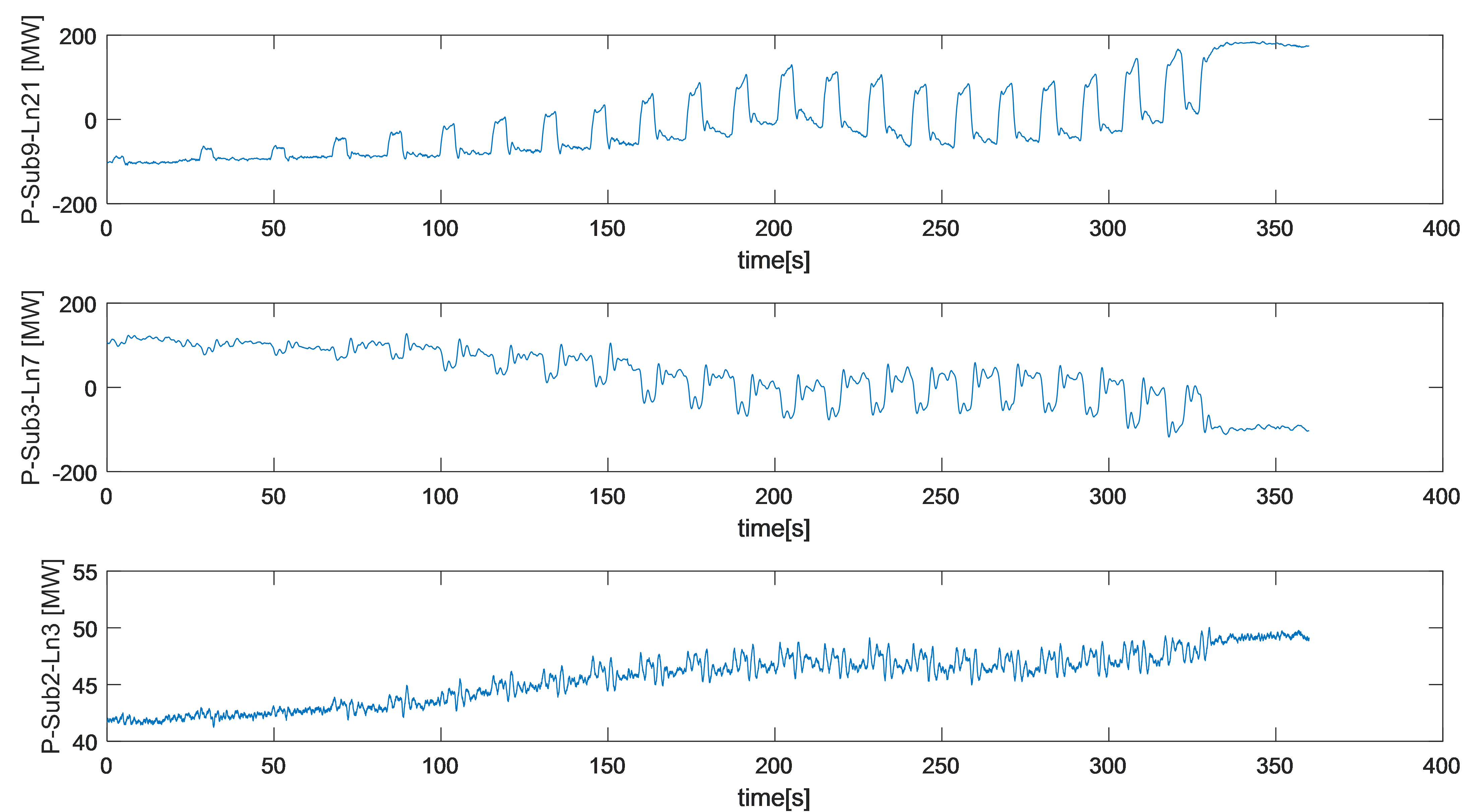}
\vspace*{-5mm}
\caption{Active Power Flow in MW during oscillation.}
\vspace*{-2mm}
\label{fig:ActivePowerFlow}
\end{figure}

On October 3, 2017, an issue in the governor of a large generator outside of the ISO-NE system created a multi-frequency process for 5 minutes. Oscillations of significant MW magnitude were observed in multiple locations of the New England power system (Case 2 of \cite{Maslennikov2018}). Fig. \ref{fig:ActivePowerFlow} shows active power flows through three selected lines indicated in Fig. \ref{fig:ISO_NE}. The non-stationary nature of the oscillation is clear, exhibiting a growing fundamental frequency and amplitude.

\subsubsection{Event ISO New England System. Proposed Methodology}

\begin{figure}[H]
\centering
\vspace*{-3mm}
\includegraphics[width = 1\columnwidth]{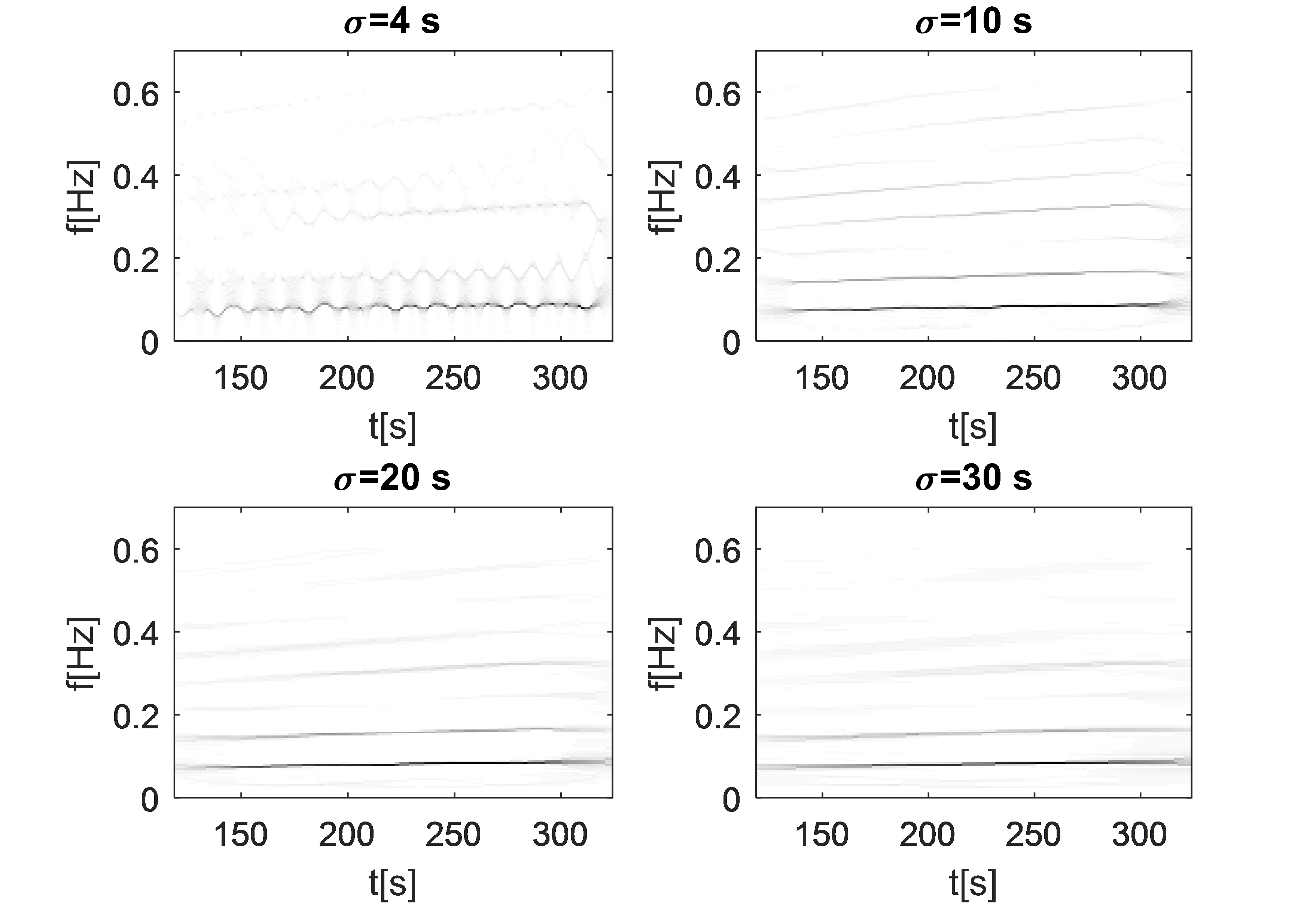}
\vspace*{-5mm}
\caption{FSST of active power flow of line Ln21 for different values of $\sigma$.}
\vspace*{-2mm}
\label{fig:Sigma_comp_SST}
\end{figure}

\begin{figure}[H]
\centering
\vspace*{-2mm}
\includegraphics[width = 0.8\columnwidth]{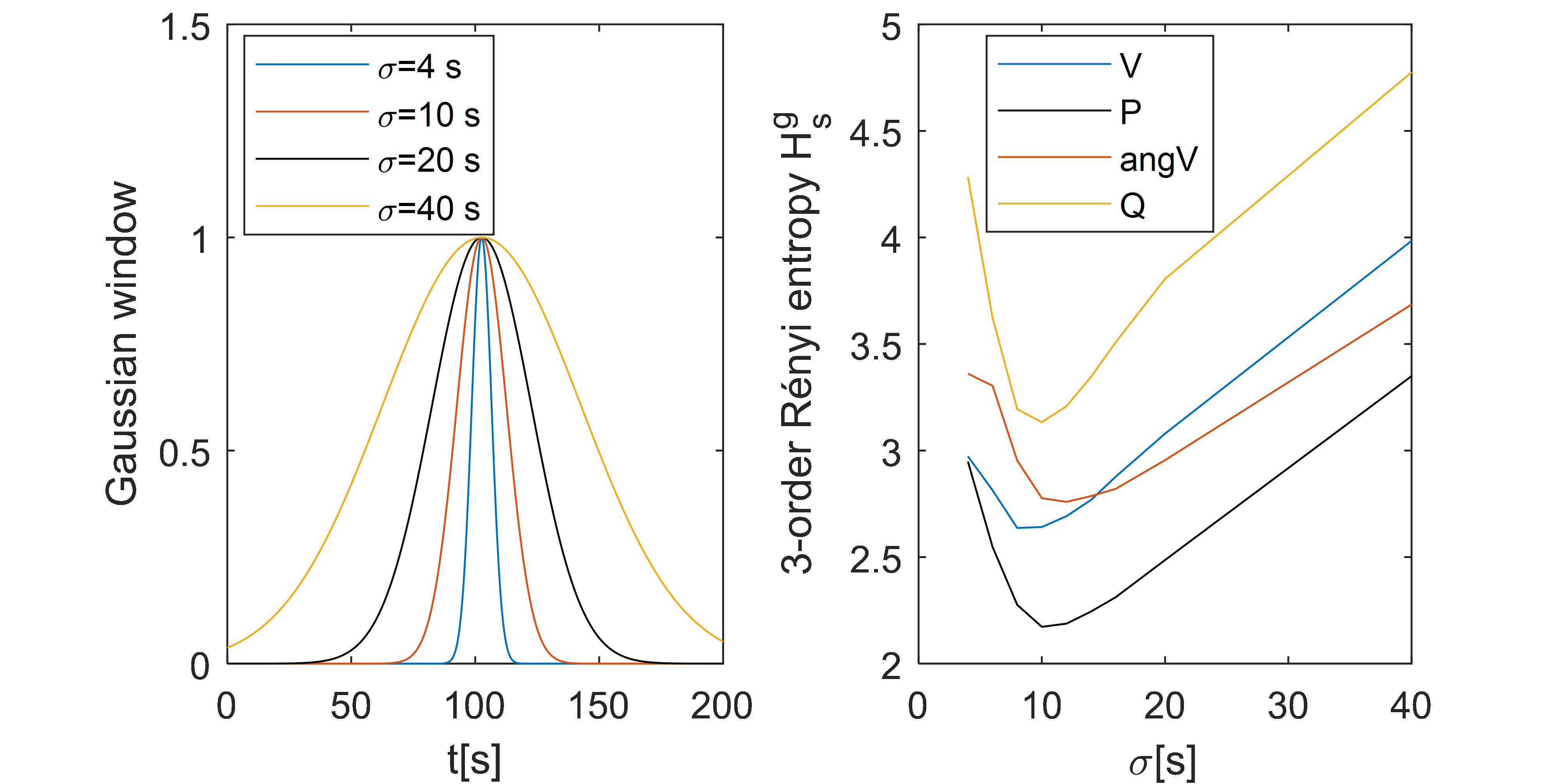}
\vspace*{-3mm}
\caption{Gaussian window and Rényi Entropy of FSST of magnitudes of line Ln21 as function of window standard deviation $\sigma$}
\vspace*{-2mm}
\label{fig:ventana_l21}
\end{figure}

Fig. \ref{fig:Sigma_comp_SST} shows FSST of active power flow of line Ln21 for different values of window standard deviation $\sigma$. Fig. \ref{fig:ventana_l21} shows that $\sigma$=10 s provides the minimum Rényi entropy of the magnitudes of line Ln21.

\begin{figure}[h]
\centering
\vspace*{-2mm}
\includegraphics[width = 0.95\columnwidth]{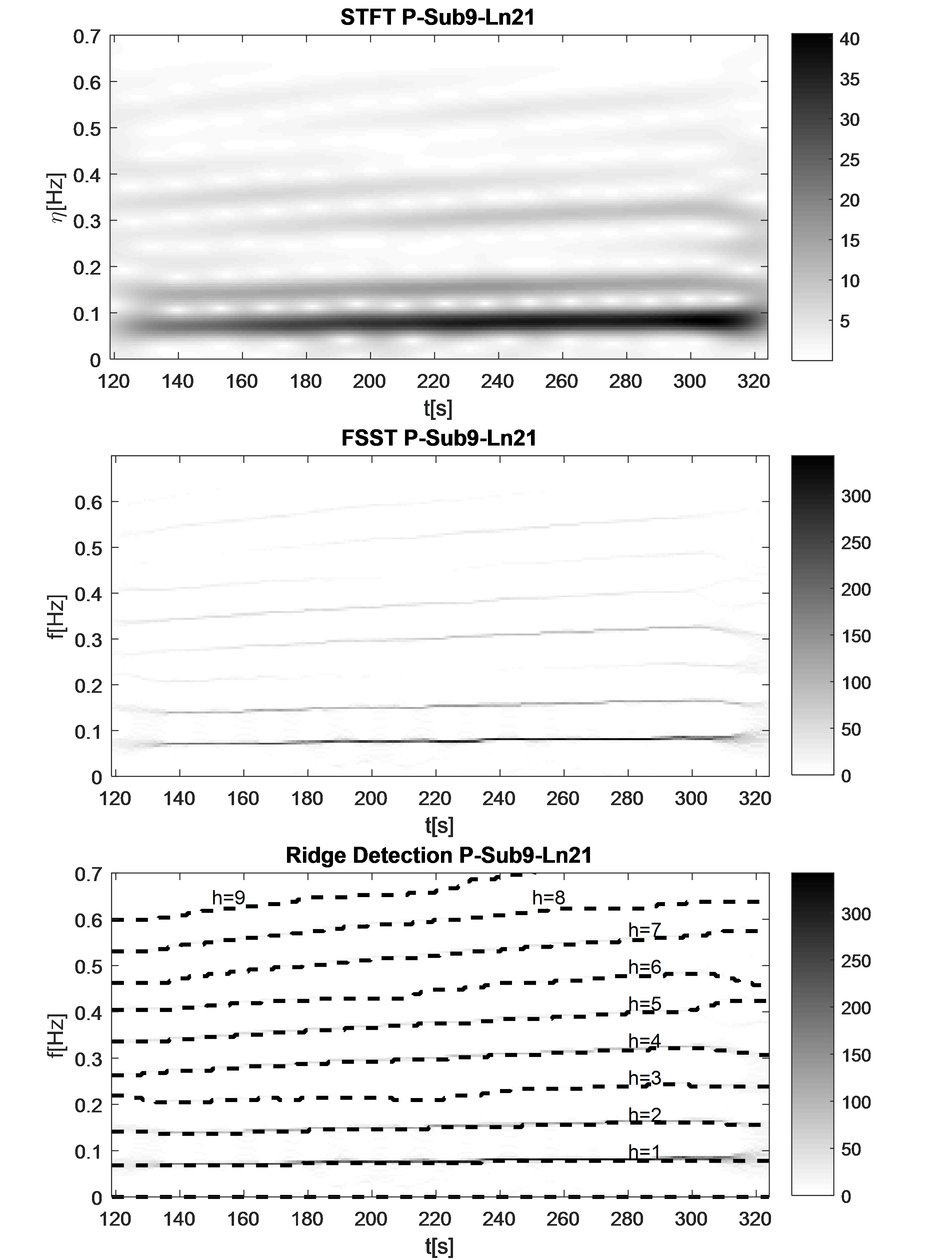}
\vspace*{-2mm}
\caption{Results for the first steps of the proposed methodology applied to the Active Power Flow through Ln21 using Gaussian window $\sigma$=10 s. Top: STFT; middle: FSST; bottom: Ridge identification.}
\vspace*{-2mm}
\label{fig:Specs1} 
\end{figure}

\begin{figure}[h]
\vspace*{-3mm}
\includegraphics[width= 1.08\columnwidth]{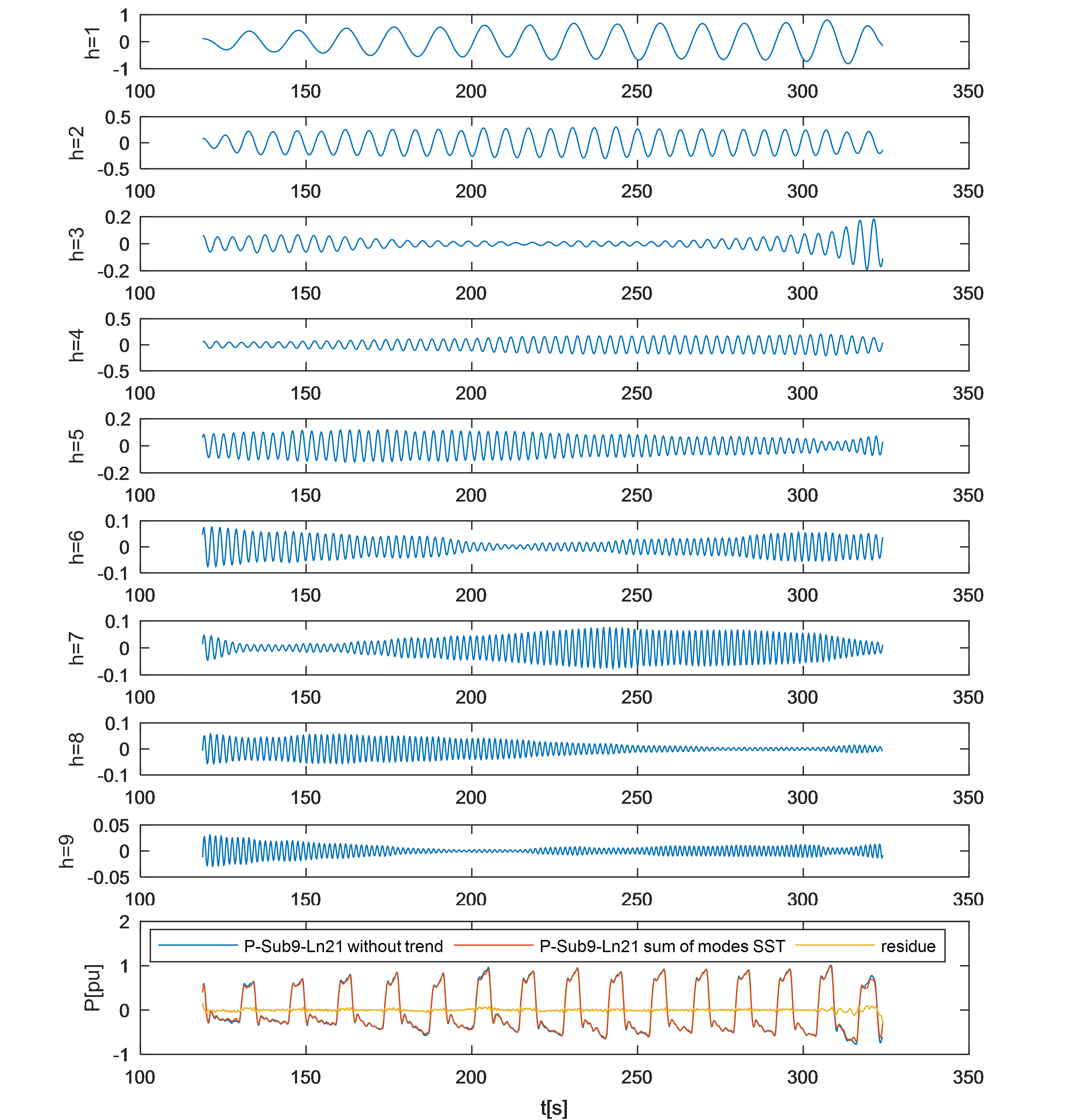}
\vspace*{-5mm}
\caption{Extracted components of Active Power Flow of Sub9-Ln21.}
\vspace*{-3mm}
\label{fig:modes}
\end{figure}

Fig. \ref{fig:Specs1} shows the STFT of the active power flow through Ln21 and the corresponding FSST with $\sigma$=10 s, as well as the results of ridge identification algorithm. Fig. \ref{fig:modes} shows the extracted components resulting from integration around each ridge. The red curve in Fig. \ref{fig:modes} is the sum of all the extracted harmonics. The residue contains mainly higher frequencies, associated to high order harmonics that were not included. The same procedure is applied to the active power $P$, voltage magnitude $V$, voltage angle $\theta$, and reactive power flow $Q$ in each of the lines. Once the decomposition of each signal is performed, the DEF method is applied to each non-stationary harmonic using \eqref{eq:flow_disip_energy}.

\begin{figure}[H]
\centering
\vspace*{-2mm}
\includegraphics[width= 1\columnwidth]{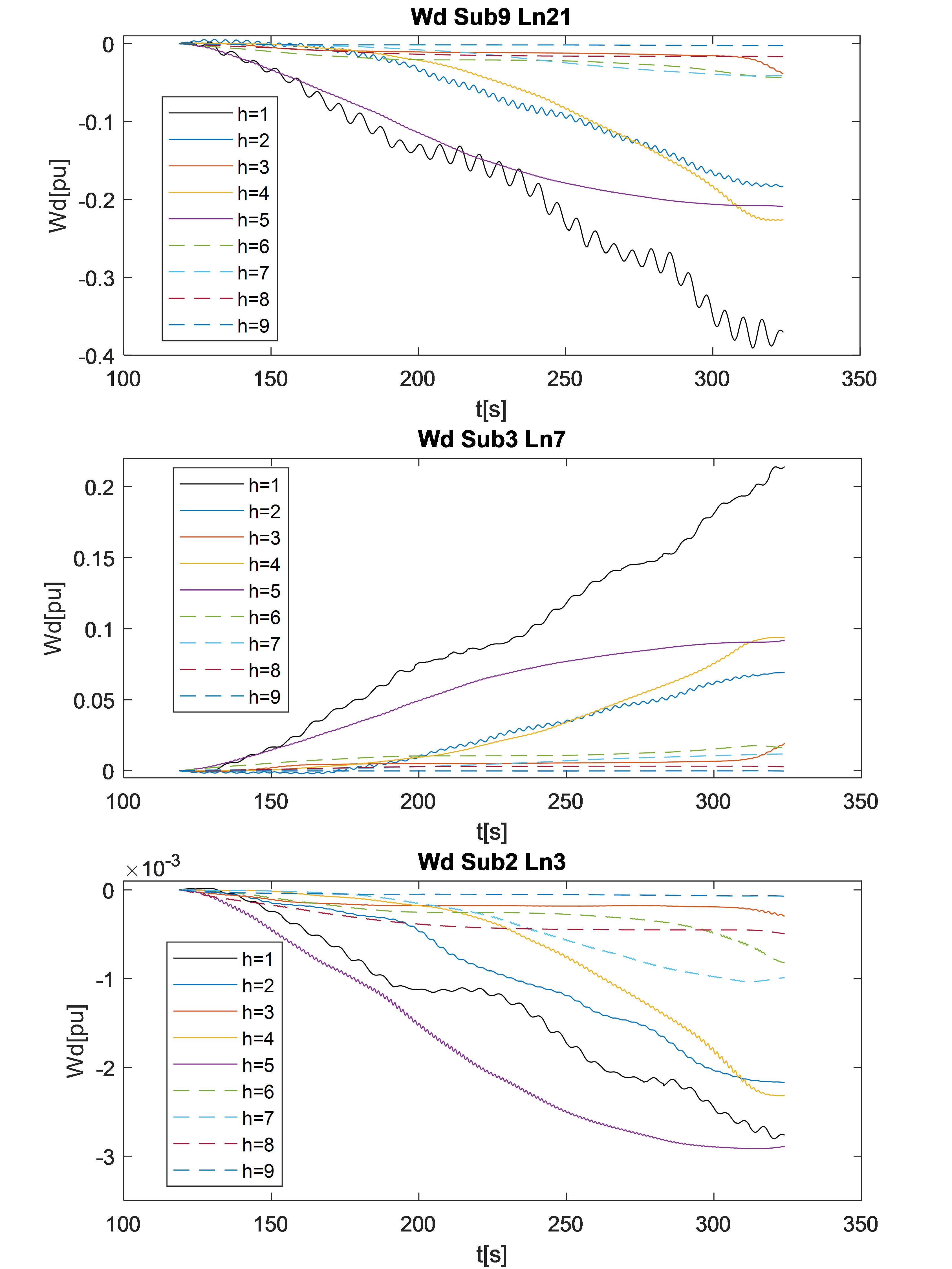}
\vspace*{-5mm}
\caption{DEF applied over non-stationary harmonics.Sign and magnitude of the rate of change of DEF allow tracing the source of oscillations.}
\vspace*{-3mm}
\label{fig:DEF}
\end{figure}

Fig. \ref{fig:DEF} shows DEF in the lines Sub9-Ln21, Sub3-Ln7 and Sub2-Ln3 with the convention indicated in Fig. \ref{fig:ISO_NE}. It can be seen that apart from the fundamental frequency of the oscillation ($h=1$), the fifth harmonic ($h=5$) has a considerable DEF because it is close to a natural system mode. Negative rate of change of DEF in Sub9-Ln21 confirms that the source of oscillation is in Area 3, as indicated in \cite{Maslennikov2018}.

\subsubsection{Event ISO New England System. Windowed DFT and Multiple Band Pass Filtering}

\begin{figure}[h]
\vspace*{-3mm}
\includegraphics[width= 1\columnwidth]{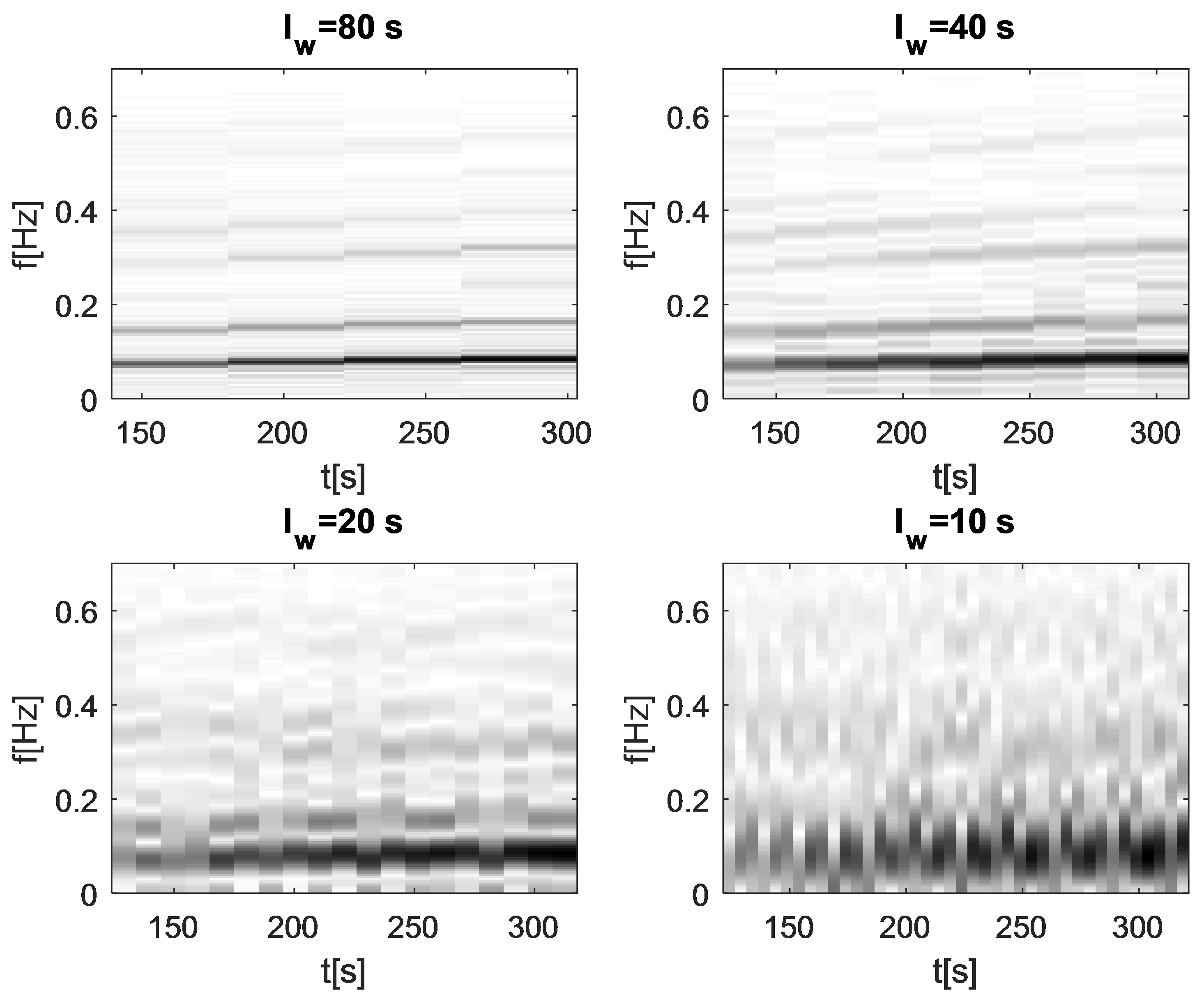}
\vspace*{-5mm}
\caption{Windowed DFT of Active Power Flow of Sub9-Ln21 for different rectangular window lengths l$_w$, with half window overlap.}
\vspace*{-2mm}
\label{fig:DFT_lw_ISONE}
\end{figure}

\begin{figure}[h]
\centering
\vspace*{-3mm}
\includegraphics[width= 0.7\columnwidth]{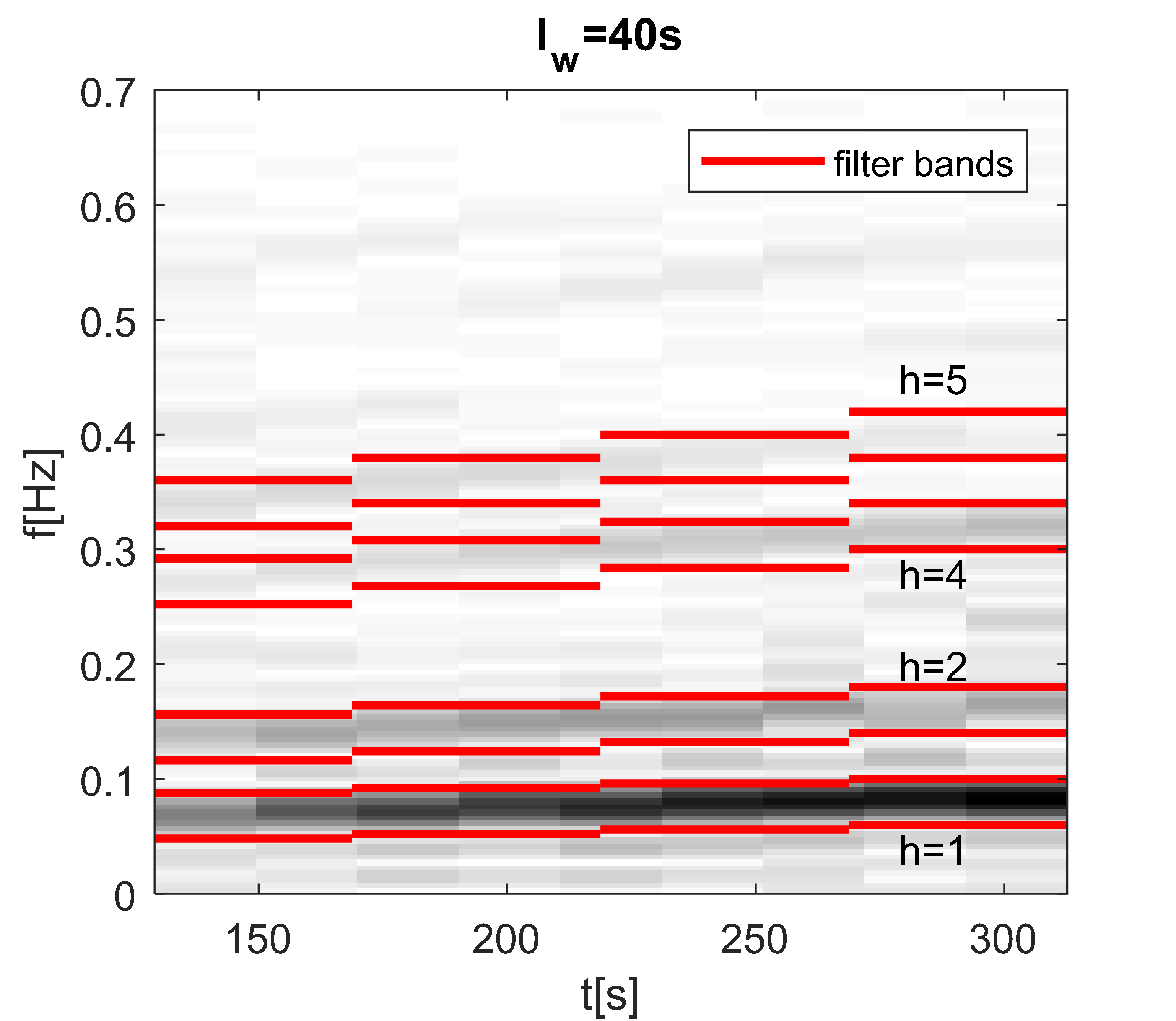}
\vspace*{-2mm}
\caption{Filtering bands defined from widowed DFT spectrogram with rectangular window of length l$_w$=40 s, with half window overlap.}
\vspace*{-2mm}
\label{fig:banda_filtros}
\end{figure}

\begin{figure}[h]
\centering
\vspace*{-2mm}
\includegraphics[width= 1\columnwidth]{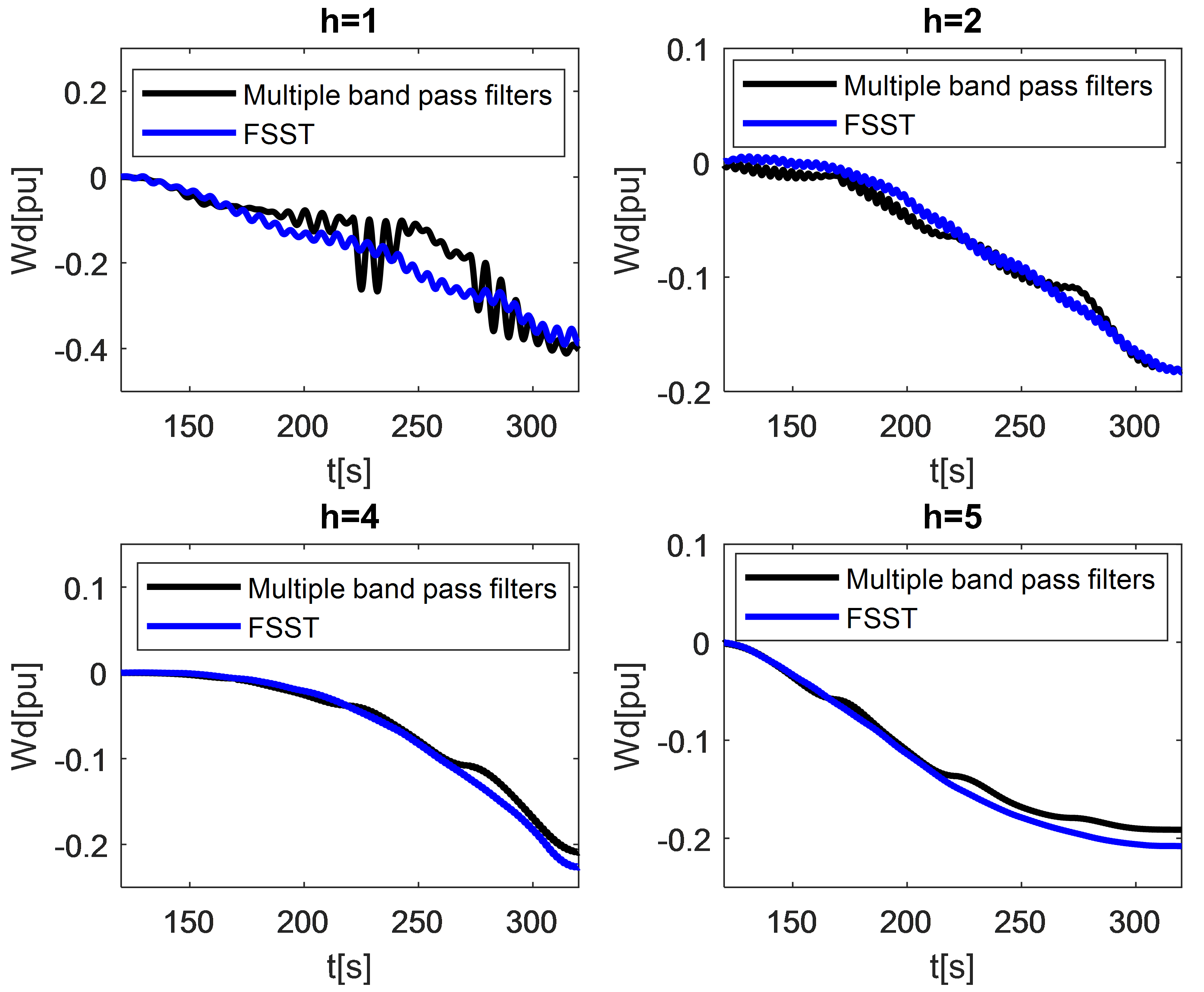}
\vspace*{-5mm}
\caption{Comparison of DEF in Line Ln21 obtained with multiple band pass filtering and FSST decomposition, for main harmonic components h.}
\vspace*{-2mm}
\label{fig:DEF_comp_var_ISONE}
\end{figure}

Having shown in the simulated case that direct application of the DEF method is unsuitable for non-stationary signals, we directly analyzed the extended method for non-stationary signals by applying windowed DFT and multiple band pass filters. Fig. \ref{fig:DFT_lw_ISONE} shows spectrogram of active power flow through line Ln21 using DFT with rectangular sliding window of length l$_w$ with half window overlap. The resulting spectrogram with a window of length l$_w$=40 s is very similar to the resulting STFT spectrum with Gaussian window with $\sigma$=10 s, shown in Fig. \ref{fig:Specs1}.
It is proposed to split the signals into 50 second intervals, to which different band pass filters are applied. The filtering bands for each of these filters are adopted from the identified spectrum, as shown in Fig. \ref{fig:banda_filtros}. Filters are considered Butterworth type with  1 dB of ripple allowed and 15 dB attenuation at both sides of the passband.
The resulting DEFs for line Ln21 are shown in Fig.  \ref{fig:DEF_comp_var_ISONE}.
As in the simulated case, the results obtained with multiple band pass filters are very similar to those obtained with the proposed methodology. 
DEF computed from the proposed methodology based on FSST presents a smoother and more monotonous variation.

\section{Conclusions}

This paper presents a methodology based on the FSST and the DEF method to analyze and locate the source of non-stationary FOs. 
The FSTT provides a TF spectrum concentrated around the ridge curves, which represent the time-varying frequencies of the oscillatory components of the signal. Component filtering is performed directly from the TF plane by integrating FSST spectrum around the identified ridge curves, without the need to carefully design multiple band pass filters. Then, DEF calculated for each non-stationary component is used to trace the source of FOs.
The results, on simulated as well as on real PMU data, show the efficiency of the proposed methodology, which is a systematic alternative method to traditional analysis with windowed DFT and band-pass filtering for the analysis of non-stationary FOs.

\appendices
\section{List of Acronyms}

\begin{acronym}[NA-MEMD] 
\acro{CWT}{continuous wavelet transform}
\acro{DEF}{dissipating energy flow}
\acro{DFT}{discrete Fourier transform}
\acro{EMD}{empirical mode decomposition}
\acro{FO}{forced oscillation}
\acro{FSST}{short-time Fourier transform based synchrosqueezing transform}
\acro{IF}{instantaneous frequency}
\acro{ISO-NE}{Independent System Operator New England}
\acro{NA-MEMD}{noise-assisted multivariate empirical mode decomposition}
\acro{PMU}{phasor measurement unit}
\acro{SST}{synchrosqueezing transform}
\acro{STFT}{short-time Fourier transform}
\acro{TF}{time-frequency}
\acro{TFR}{time-frequency reassignment}
\acro{WECC}{Western Electricity Coordinating Council}
\acro{WSST}{continuous wavelet transform based synchrosqueezing transform}
\end{acronym}





\bibliographystyle{IEEEtran}
\bibliography{biblio}
\end{document}